\documentclass[journal]{IEEEtran}
\IEEEoverridecommandlockouts
\usepackage{cite}

\usepackage{amsmath, amssymb, amsfonts, mathtools, bm}
\usepackage{amsthm}
\usepackage{algorithm}
\usepackage{algorithmic}
\usepackage{graphicx}
\usepackage{textcomp}
\usepackage{xcolor, float}
\usepackage{tabularx}
\usepackage{textcomp}
\usepackage{diagbox}
\usepackage{svg}
\usepackage{booktabs}
\usepackage{subfigure}
\usepackage{setspace}
\DeclareUnicodeCharacter{2061}{}

\allowdisplaybreaks
\renewcommand{\baselinestretch}{1}

\begin{document}

\title{Knowledge-Driven Resource Allocation for D2D Networks: A WMMSE Unrolled Graph Neural Network Approach}
\author{
    Hao~Yang,~\IEEEmembership{Student~Member,~IEEE,}
    Nan~Cheng,~\IEEEmembership{Member,~IEEE,}
    Ruijin~Sun,~\IEEEmembership{Member,~IEEE,}
    Wei~Quan,~\IEEEmembership{Member,~IEEE,}
    Rong~Chai,~\IEEEmembership{Senior~Member,~IEEE,}
    Khalid~Aldubaikhy,~\IEEEmembership{Member,~IEEE,}
    Abdullah~Alqasir,~\IEEEmembership{Member,~IEEE,}    and~Xuemin (Sherman) Shen,~\IEEEmembership{Fellow,~IEEE}

\thanks{
\par Hao Yang, Nan Cheng, and Ruijin Sun are with the State Key Lab. of ISN and School of Telecommunications Engineering, Xidian University, Xi’an 710071, China (e-mail: yh\_xd210@xidian.edu.cn; dr.nan.cheng@ieee.org; sunruijin@xidian.edu.cn).
\par Wei Quan is with School of Electronic and Information Engineering, Beijing 
Jiaotong University, Beijing 100044, China (e-mail: weiquan@bjtu.edu.cn).
\par Rong Chai is with School of Communication and Information Engineering, Chongqing University of Posts and Telecommunications, Chongqing 400065, China (e-mail: chairong@cqupt.edu.cn).
\par K. Aldubaikhy and A. Alqasir are with the Department of Electrical Engineering, College of Engineering, Qassim University, Qassim, Saudi Arabia
(e-mail: \{khalid, a.alqasir\}@qec.edu.sa).
\par Xuemin (Sherman) Shen is with the Department of Electrical and Computer Engineering, University of Waterloo, Waterloo, N2L 3G1, Canada (e-mail: sshen@uwaterloo.ca).
\par \emph{Corresponding Author: Ruijin Sun.}
}

}

\maketitle

\IEEEdisplaynontitleabstractindextext

\IEEEpeerreviewmaketitle

\begin{abstract}
This paper proposes an novel knowledge-driven approach for resource allocation in device-to-device (D2D) networks using a graph neural network (GNN) architecture. To meet the millisecond-level timeliness and scalability required for the dynamic network environment, our proposed approach incorporates the deep unrolling of the weighted minimum mean square error (WMMSE) algorithm, referred to as domain knowledge, into GNN, thereby reducing computational delay and sample complexity while adapting to various data distributions. Specifically, the aggregation and update functions in the GNN architecture are designed by utilizing the summation and power calculation components of the WMMSE algorithm, which leads to improved model generalization and interpretabiliy. Theoretical analysis of the proposed approach reveals its capability to simplify intricate end-to-end mappings and diminish the model exploration space, resulting in increased network expressiveness and enhanced optimization performance. Simulation results demonstrate the robustness, scalability, and strong performance of the proposed knowledge-driven resource allocation approach across diverse communication topologies without retraining. Our findings contribute to the development of efficient and scalable wireless resource management solutions for distributed and dynamic networks with strict latency requirements.
\end{abstract}


\section{Introduction}
In the era of 6G, mobile communication networks are envisioned to provide a wide variety of services and applications, from data-intensive services such as extended reality (XR), reliable and low-latency services such as autonomous driving and remote surgery, to the soaring intelligent services such as metaverse and ChatGPT \cite{Zhou2019GLOBECOM}. Furthermore, 6G networks are becoming increasingly complex and dynamic as the emergence and fast development of space-air-ground integrated networks (SAGINs) significantly enlarge the network scale and require efficient management of multi-dimensional resources\cite{CHENG20221}. The complexity poses a significant challenge to wireless network management, such as resource allocation schemes and task scheduling, to fulfill the service requirements, especially delay-sensitive and reliability services, where a fault or delayed decision may lead to fatal outcomes. Therefore, it is critical to design efficient, responsive, and scalable wireless network management schemes in 6G networks.

Wireless resource allocation plays a pivotal role in network management to allocate spatial and temporal wireless resources for certain goals, such as maximizing the transmission rate or minimizing the transmission delay or energy consumption. A plethora of model-based iterative algorithms, such as iterative water-filling type algorithms \cite{r1}, WMMSE algorithms \cite{r2}, and successive convex approximation algorithms \cite{r3,Sun2023}, have been proposed based on the convex optimization theory to address the wireless resource allocation problem. These algorithms have successfully solved classical resource allocation problems, commonly with small network scale and low-level network dynamics. However, as the network scale increases, the high computational complexity associated with multiple iterations of these algorithms can hardly meet the stringent milliseconds-level service requirements.

Due to the efficient real-time computational capabilities, deep learning techniques have found application in diverse areas of wireless communication systems\cite{Wireless_DL}, such as UAV-assisted IoT applications\cite{cheng2023ai},spectrum Sharing\cite{li2022edge}, resource management\cite{peng2020deep}, and mobile computing offloading\cite{cheng2019space}. Ye {\it et al.} employed deep neural networks (DNNs) for channel estimation and signal detection, achieving efficient handling of channel distortion\cite{8052521}. He \textit{et al.} investigated the utilization of convolutional neural network (CNN)-based architectures for channel estimation in beamspace millimeter-wave massive multiple-input-multiple-output (MIMO) systems, surpassing the most advanced compressed sensing-based algorithms. Tang and Wong utilized bidirectional long short-term memory (LSTM) in mobile edge computing systems to handle user scheduling problems in ad hoc networks, effectively minimizing the average age of information \cite{9253665}.
To solve the resource allocation problem for weighted sum rate maximization, multi-layer perceptrons (MLP) \cite{r4} and  CNN \cite{r5} are employed to approximate the WMMSE algorithm, using outputs of algorithms as labels and reducing computational complexity. In \cite{r6}, the objective function, i.e., the weighted sum rate, is regarded as the loss function, achieving better performance. These studies presented exhibit impressive performance and low inference complexity. However, applying them effectively to radio resource management for an arbitrary number of users faces a significant challenge. Commonly utilized neural networks, including DNN, CNNs, recurrent neural networks, and attention-based Transformer models, are not readily scalable for an arbitrary number of users, as their input and output dimensions must remain constant. Therefore, designing scalable neural network architectures is crucial for effectively managing wireless resources, given the dynamic fluctuations in user numbers within mobile applications.

In this context, a graph structure with expandable connecting nodes is more suitable for capturing the dynamic characteristics of wireless networks. By incorporating such graph structure into neural networks, GNNs are envisioned as a potential solution to realize scalable resource allocation\cite{chen2021gnn,wang2022joint}. A random edge graph neural network (REGNN) is proposed to enhance scalability and generalization for optimal power control in interference channels \cite{r7}. Addressing the limitations of REGNN in heterogeneous agents and multi-antenna systems, the interference graph convolutional network (IGCNet) is proposed in \cite{r8}. Furthermore, a message-passing graph neural network (MPGNN) is presented for tackling large-scale wireless resource management problems, such as beamforming, user association, and channel estimation\cite{r9}. The authors established the equivalence between MPGNN and distributed optimization algorithms, showcasing its performance and generalization capabilities. While GNNs demonstrate scalability, their intrinsic learning approach primarily relies on statistical distributions with poor interpretability, leading to struggles to accommodate varying distributions and necessitating a large amount of training data for a particular distribution.
In the context of radio resource management, collecting identical distribution training data is time-consuming and costly, and the dynamic nature of wireless networks causes a dataset shift that degrades model performance. 

Unlike deep learning, model-based iterative algorithms can consistently achieve solutions with theoretical performance guarantees. Integrating domain knowledge in model-based algorithms and neural networks, known as knowledge-driven methods, can simplify the architecture of machine learning systems, decrease training overhead, enhance the interpretability of decisions, and increase their practical utility \cite{von2021informed}. Deep unrolling\cite{monga2021algorithm} provides an effective solution for integrating domain knowledge with iterative algorithms. The main idea is to design neural networks by leveraging the structure of classical iterative algorithms, incorporating the iterative structure of the algorithm into each layer of the network. This approach treats network layer as an iteration in the original iterative optimization algorithm and learns the network parameters from the data. Deep unrolling combines the benefits of data-driven learning with the domain knowledge embedded in the iterative algorithm, resulting in improved performance and generalization capabilities. In the field of image processing, deep unrolling has successfully addressed several challenging problems, including image restoration\cite{r10}, deep image deblurring\cite{r11}, and image super-resolution\cite{r12}.
In the field of wireless communications, deep unrolling projected gradient descent (PGD) algorithm into a neural network has shown better accuracy with lower and more flexible computational complexity in MIMO detection problems \cite{samuel2019learning}. A low-complexity deep neural network-based MIMO detector was proposed using the multipliers algorithm's deep unrolling alternating direction approach\cite{un2019deep}. In \cite{shi2021algorithm}, the original iterative shrinkage thresholding algorithm is transformed into an unrolled RNN, maintaining the robustness of the algorithm and improving estimation accuracy. The iterative WMMSE algorithm is unrolled into a layer-by-layer CNN structure, introducing trainable parameters to replace the high-complexity operations in forward propagation, reducing computational complexity for efficient performance and enhancing neural network generalization \cite{r16}. In addition to unrolling model-based iterative algorithms such as DNN and RNN, unrolled GNN has the advantages of scalability and interpretability. A deep unrolling architecture based on GNN is proposed in \cite{r13}, which only learns key parameters of the WMMSE algorithm with GNN without unrolling iterations in WMMSE algorithm as layers in GNN. As demonstrated in \cite{r15}, the alignment of the GNN architecture with the algorithm may potentially enhance the representation of GNN and thus increase the sample complexity. Therefore, the investigation of GNN unrolling with accurate alignment between GNN layers and algorithm iterations is required to improve the performance further.

In this paper, we propose a novel knowledge-driven GNN architecture-based resource allocation approach for D2D networks, guided by the WMMSE algorithm's unrolling, aiming to manage radio resources in D2D networks. 
Specifically, to align one iteration of the WMMSE algorithm with a one layer GNN, message passing and aggregation functions are designed based on the unrolling WMMSE algorithm, which utilize two cascaded GNN modules respectively for hierarchical feature extraction and node update. By adopting the structure and domain knowledge of the WMMSE algorithm, our proposed approach retains the algorithm's robustness while effectively reducing its computational delay, decreasing the sample complexity of the neural network, and adapting to various data distributions. The main contributions of the paper can be summarized as follows.
\begin{enumerate}
    \item We propose a novel GNN architecture guided by a deep unrolling WMMSE algorithm, named UWGNN. This approach leverages the summation part of the WMMSE algorithm to design the aggregation function in GNN while adopts the power calculation formula of a single node to design the update function of GNN. 
    \item We conduct a theoretical analysis to demonstrate the validity of deep unrolling in the UWGNN, where layers in GNN are alignment with iterations in WMMSE algorithm. The deep unrolling model is scrutinized from two aspects: network mapping and the size of network exploration space. Through this analysis, it is concluded that deep unrolling make the mapping relationship of the model more accurate and reduce the exploration space of the model.
    \item Our proposed UWGNN shows strong performance, robustness, and scalability through extensive simulations. Moreover, the architecture exhibits excellent generalization performance when dealing with diverse data distributions and communication topologies without necessitating retraining. Experiments show that the approach enables on-demand reduction of delay in communication network resource allocation. 
\end{enumerate}
The remainder of this paper is organized as follows. Section II introduces the communication model and the formulae for resource allocation problems. In Section III, we present the WMMSE algorithm along with our proposed unrolling network architecture. In Section IV, we propose a theoretical hypothesis for the validity of the unrolling technique. In Section V, we demonstrate the effectiveness of our proposed approach through numerical experiments. Finally, Section VI summarizes and concludes the paper.

\section{System Model and Problem Formulation}

\subsection{System Model}
We consider a D2D scenario consisting of $N$ single-antenna transceiver pairs. Let $p_i$ denote the transmission power that transmitter $i$ uses to send a baseband signal $s_i$ to receiver $i$. That is, the transmission signal $x_i=\sqrt{p_i} s_i$. Then, the received signal at receiver $i$ is 
\begin{align}
 &y_i=h_{ii}x_i+\sum_{j=1,j\neq i}^{N}h_{ij}x_j+n_i,\: \forall i,
\label{equ.2-1}
\end{align}
where $h_{ii}\in \mathbb{R}$ represents direct channel between transmitter $i$ and receivers $i$, $h_{ij} \in \mathbb{R}$ with $i \neq j$ interference channel from transmitter $j$ to receiver $i$, and $n_i\in \mathbb{R}$ denotes the additive noise following the complex Gaussian distribution $\mathcal{CN}(0, \sigma^2)$. Based on the receive threshold equalization $u_i$, the signal recovered by receiver $i$ can be obtained as $\hat{s}_i=u_i y_i$. Assuming that the signals of different users are independent of each other and receiver noises, the signal-to-interference-plus-noise ratio (SINR) of receiver $i$ is expressed as
\begin{equation}
SINR_i=\frac{|h_{ii}|^2 p_i}{\sum_{j\neq i}^{N}|h_{ij}|^2 p_j+\sigma ^2},\: \forall i, 
\label{equ.2-2}
\end{equation}
where $p_i$ represents the power of
transmitter $i$, and $0\leq p_i\leq p_\mathrm{max}$ with $p_\mathrm{max}$ denoting as the max transmission power of the transmitter. 

Our objective is to maximize the weighted sum rate by optimizing the transmission power, formulated as
\begin{align}
\centering
\max_{\mathbf{p}} ⁡&\sum_{i=1}^{N} \lambda_i\mathrm{log_2}\left (1+\frac{|h_{ii}|^2 p_i}{\sum_{j\neq i}^{N}|h_{ij}|^2 p_j+\sigma ^2}\right),\label{equ.2-3} \\
\mathrm{s. t.} &\; 0\leq p_i\leq p_\mathrm{max},\; \forall i,\notag
\end{align}
where the weight $\lambda_i$ represents the priority of transmitter $i$ in the sum rate problem, and the power vector is expressed as $\mathbf{p}=[p_1,\dots,p_N]$.

\subsection{WMMSE Algorithm}
The non-convex nature of Problem (\ref{equ.2-3}) arises from the objective function. Many iterative algorithms have been proposed to solve it effectively, of which the WMMSE algorithm\cite{r2} is the most classical one. The main idea of the algorithm is to equivalently transform the weighted sum-rate maximization problem into a problem of minimizing a weighted sum of mean squared errors (MSE), as follows: 
\begin{align}
\centering
⁡\min_{\mathbf{u,v,w} } &\sum_{i=1}^{N}\lambda_i(w_i e_i-\mathrm{log}\: w_i) \notag\\ 
\mathrm{s. t.} \; &0\leq{v_i}^2\leq p_\mathrm{max},\; \forall i, \label{equ.2-6}
\end{align}
where $w_i\ge 0$ is an introduced auxiliary variables indicating the weight for transmitter $i$. $v_i=\sqrt{p_i}$, and $\mathbf{u}=[u_1,\dots,u_N], \mathbf{v}=[v_1,\dots,v_N], \mathbf{w} =[w_1,\dots,w_N]$, respectively. $e_{i}\triangleq \mathbb{E}_{s, n}\left[\left(\hat{s}_{i}-s_{i}\right)^2\right]$ is the MSE covariance of the transmission signal $s_i$ and the recovery signal $\hat{s}_i$, represented as 
\begin{align}
\centering
e_{i}=\left(1-u_{i}h_{ii}v_{i}\right)^2 +\sum_{j \neq i} \left({u}_{i}h_{ij}v_{j}\right)^2 +\sigma^{2}{u}_{i}^2,\; \forall i. 
\end{align}

It has been demonstrated in \cite{r2} that the WMMSE problem presented in (\ref{equ.2-6}) is equivalent to the problem of maximizing the sum-rate as depicted in (\ref{equ.2-3}), with both problems sharing an identical optimal solution denoted as $v_i$. Subsequently, the weighted sum-MSE minimization problem is decomposed into three separate optimization subproblems, each of which can be solved iteratively. Since the subproblems associated with the optimization variable vectors $\{\mathbf{u,v,w}\}$ are convex in nature, the algorithm utilizes a block coordinate descent approach to solve the WMMSE problem in (\ref{equ.2-6}). More specifically, by sequentially fixing two of the three variables $\{u_i,w_i,v_i\}$ and simultaneously updating the third variable, the WMMSE formula is as follows:
\begin{align}
\centering
 {u_i}^{(k)} &= \frac{h_{ii} {v_i}^{(k-1)}}{\sigma ^2+\sum_{j}^{} h_{ij}^2 {v_j}^{(k-1)}{v_j}^{(k-1)}},\: \forall i \label{equ.2-7} \\
 {w_i}^{(k)} &= \frac{1}{1-{u_i}^{(k)}h_{ii} {v_i}^{(k-1)}},\: \forall i \label{equ.2-8} \\
 {v_i}^{(k)} &= \frac{ \lambda_i{u_i}^{(k)}h_{ii}{w_i}^{(k)}}{\sum_{j}^{} h_{ji}^2 {u_j}^{(k)}{u_j}^{(k)}{w_j}^{(k)}},\: \forall i \label{equ.2-9}
\end{align}
where $k=1,...,K$ represents the number of iterations. The detailed WMMSE algorithm is outlined in Algorithm 1.
\begin{algorithm}
\setstretch{1.1}
	\renewcommand{\algorithmicrequire}{\textbf{Input:}}
	\renewcommand{\algorithmicensure}{\textbf{Output:}}
	\caption{WMMSE Algorithm}
	\label{alg1}
	\begin{algorithmic}[1]
		\STATE Initialization:$\left\{ {v_{i}} \right\} $ to satisfy
 $0\leq{v_i}^2\leq p_\mathrm{max}.$ The current iteration index $k = 1$. 
		\REPEAT
		\STATE Update $ u_{i}^{(k)} $ based on Equation~(\ref{equ.2-7});
		\STATE Update $w_{i}^{(k)}$ based on Equation~(\ref{equ.2-8});
		\STATE Update $v_{i}^{(k)}$ based on Equation~(\ref{equ.2-9});
 \STATE $k=k+1$;
		\UNTIL the convergence condition is met,
		\ENSURE transmission power $p_i=v_i^2$.
	\end{algorithmic} 
\end{algorithm}

Although the WMMSE algorithm has demonstrated high performance in various wireless communication systems, some of its shortcomings limit its practical application. Firstly, the algorithm is prone to get trapped in local optima. Additionally, the computational time required for the WMMSE algorithm to converge is significant, particularly in large-scale networks. 

\section{Neural Network Architecture Design Using WMMSE Algorithm}
To enhance online computational efficiency while preserving the interpretability of the WMMSE algorithm, we propose a knowledge-driven GNN approach for transmission power allocation in D2D networks. Our proposed method incorporates the unrolled WMMSE algorithm as the message aggregation and combination functions within the GNN. In what follows, we first briefly introduce GNN and unrolling technique, then present the knowledge-driven GNN. 

\subsection{Preliminaries}

\subsubsection{Graph Neural Networks}

GNNs were initially designed to process non-Euclidean structured graphs data \cite{first_gnn}. Unlike traditional neural networks that operate on a fixed grid of inputs, GNNs can handle data with arbitrary connectivity, making them well-suited for tasks such as node classification, graph classification, and clustering detection. Particularly, GNNs operate by iteratively passing messages between nodes in the graph, updating the node representations based on the received information from neighboring nodes. This process can be thought of as a form of message passing, where each node can aggregate information from its local neighborhood and integrate it into its representation.

The aggregation function is primarily utilized to consolidate the neighborhood features of nodes from their neighboring nodes and connected edges. In contrast, the combined function is responsible for updating the current node features based on the previous iteration node features and the neighborhood features. Formally, the aggregate and combine rules of the $k$-th layer at node $i$ in GNNs are respectively expressed as
\begin{align}
\centering
\boldsymbol{\alpha}_{i}^{(k)} & =\operatorname{AGGREGATE}^{(k)}\left(\left\{\boldsymbol{\beta}_{j}^{(k-1)}: j \in \mathcal{N}(i)\right\}\right), \\
\boldsymbol{\beta}_{i}^{(k)} & =\operatorname{COMBINE}^{(k)}\left(\boldsymbol{\beta}_{i}^{(k-1)}, \boldsymbol{\alpha}_{i}^{(k)}\right)
\end{align}
where ${\beta}_{i}^{(k)}$ represents the feature vector of node $i$ at either the $k$-th layer or after the $k$-th iteration. $\mathcal{N}(i)$ is the set of neighbor nodes of $i$, and ${\alpha}_{i}^{(k)}$ is an intermediate variable. 

\subsubsection{Algorithm Unrolling}

Algorithm unrolling, also referred to as deep unrolling or unfolding, represents a technique that bridges the gap between deep learning and traditional iterative models, enabling the amalgamation of domain knowledge and data-driven learning. The fundamental concept of deep unrolling is transforming an iterative inference algorithm into a hierarchical structure that mimics a neural network. Each layer of the neural network corresponds to each iteration of the algorithm. 
Gregor and LeCun proposed deep unrolling seminal work \cite{gregor2010learning}, which has been used to connect various iterative algorithms, such as those used in sparse coding, to diverse neural network architectures. It is possible to unfold an N-step iterative inference algorithm into an N-layer neural network with trainable parameters. This aims to enhance the model performance by leveraging a computationally-lighter neural network. 

Unrolled networks boast high parameter efficiency and require less training data than popular neural networks. Therefore, this approach efficiently counters the lack of interpretability normally found in traditional neural networks. This approach provides a systematic link between traditional iterative algorithms and deep neural networks, leading to efficient, interpretable, and high-performance network architectures.

\subsection{Graph Representation of the Weighted Sum Rate Maximization Problem}
\begin{figure}[htbp]
\centering
\includegraphics[width=0.4\textwidth]{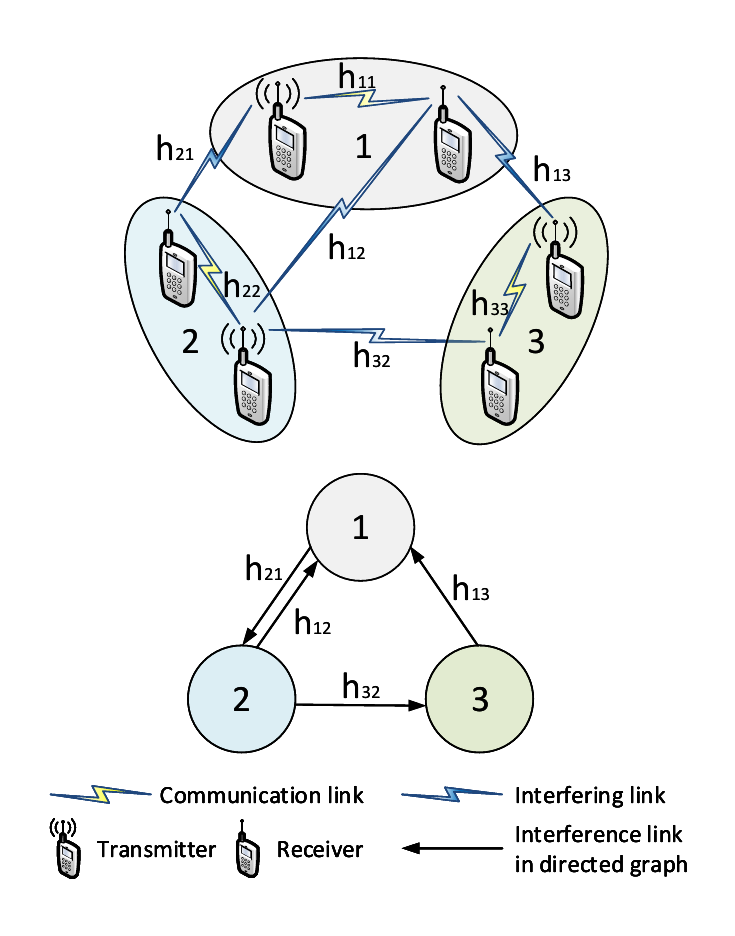}
\caption{Graph modeling of the D2D communication network. We consider a pair of D2D users as a node in the graph and the interference link between transmitter $j$ and receiver $i$ as an edge between node $j$ and node $i$.}
\label{Fig.0}
\end{figure}
\begin{figure*}[htbp]
\centering
\includegraphics[width=0.8\textwidth]{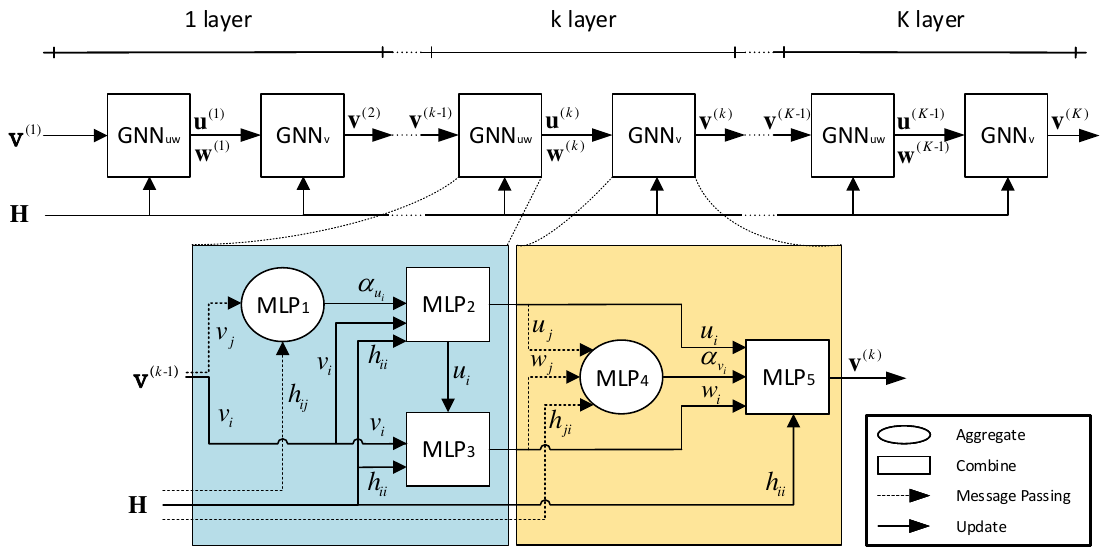}
\caption{Knowledge-driven graph neural network architecture. The proposed GNN model utilizes a two-module architecture inspired by the unrolling WMMSE algorithm. The first GNN module focuses on calculating $u_i$ and $w_i$, while the second module computes node power $v_i$. This hierarchical feature extraction and node update strategy allows for deep learning of data features and more effective cognition within the GNN architecture.}
\label{Fig.1}
\end{figure*}
Before presenting the knowledge-driven neural network architecture, we considered the D2D network as a directed graph with node and edge features. As shown in Fig. \ref{Fig.0}, we consider a pair of D2D communication users as a node in the graph and the interference link between transmitter $j$ and receiver $i$ as an edge between node $j$ and node $i$. Typically, node features include attributes such as node labels, node degrees, and node positions, while edge features may include attributes such as edge weights, edge types, and edge directions. For problem (\ref{equ.2-6}), node features contain the weighted factor $\lambda_i$, channel gain $h_{ii}$ between D2D pair, transmission power $v_i$ of transmitter$i$, and resource allocation intermediate variables $u_i, w_i$, etc, whereas edge feature includes the channel gain of the interfering channel $h_{ij}$.

The modeled wireless channel directed graph is mathematically represented as $G=(V,E)$, where $V$ is the set of nodes, $E$ is the set of edges. We denote the feature of node $i$ by notation vector $\mathbf{z}_i$, represented as
\begin{equation}
\mathbf{z}_i=[ \lambda_i,h_{ii},v_i,\mathbf{u_i,w_i}]^\top, \: \mathbf{z}_i \in \mathbb{C} ^{({3+d_u+d_w})\times 1}
\label{equ.20}
\end{equation}
where $\lambda_i,\;h_{ii},\;v_i$ are one-dimensional variables, $u_i$ and $w_i$ are a $d_{u}$-dimension vector and a $d_{w}$-dimension vector, respectively, expanded by the one-dimensional variables, to enhance the feature extraction capability of GNNs.

To improve the feature extraction capability of GNN, we propose expanding the one-dimensional variable $u_i$ and $w_i$ into a $d_{u}$-dimension vector and a $d_{w}$-dimension vector, respectively. 
The node feature matrix $\mathbf{Z}$ can be expressed as $\mathbf{Z}=\left [\mathbf{z}_1,\dots,\mathbf{z}_N\right]$. 
The edges adjacency feature matrix $\mathbf{A}\in \mathbb{C} ^{{N}\times {N}}$ is given by
\begin{equation}
\mathbf{A}_{\left(i,j \right) } = \left\{\begin{matrix} 
 0 , \qquad {\mathrm{if} \left \{i,j \right \} \notin E} \\ 
 h_{ij}, \qquad{\mathrm{otherwise}} 
\end{matrix}\right.
\label{equ.21}
\end{equation}

By defining nodes feature matrix $\mathbf{Z}$ and edges feature matrix $\mathbf{A}$, the considered D2D scenario is converted as a directed graph.
Based on this, we will develop an effective algorithmic knowledge inspired graph neural network architecture.

\subsection{Proposed Knowledge-Driven GNN Architecture}

Although the non-Euclidean data structure of GNN has the advantage of handling communication topological information compared to other deep learning models, its internal structure design is also essential. The aggregate and combined functions of most GNNs are typically designed with consideration of graph data structures and features, such as permutation invariance and self-attentiveness mechanisms, etc. Although these data-driven design approaches have excellent end-to-end nonlinear mapping performance, they are usually interpretable to a limited extent as black box models.

Inspired by the deep unrolling technique, we propose a novel GNN architecture based on the unrolling WMMSE algorithm, named UWGNN.
Specifically, in the message passing process of the GNN, we design the message passing and aggregation functions by utilizing the sum operation of neighborhood information in the denominators of (\ref{equ.2-7}) and (\ref{equ.2-9}) in the WMMSE algorithm. Instead of aggregating all node and edge features into the aggregation function at once, we selectively input the features of the aggregation function based on the sum operation. Typically, a single-layer GNN involves only one round of message passing and aggregation. In contrast, as shown in (\ref{equ.2-7}) and (\ref{equ.2-9}), the WMMSE algorithm requires two rounds of aggregation for neighbor messages within one cycle, and the information dimensions of the two processes are different. Therefore, we design two different aggregation functions corresponding to two GNN modules in cascade. For the node feature update, we design three update steps based on the WMMSE algorithm, corresponding to the updates of $u_i$, $w_i$, and $v_i$. Similar to the idea of the algorithm to fix one variable and update the remaining variables, our GNN architecture first updates $u_i$, then updates $w_i$, and finally passes the new $u_i$ and $w_i$ to updates $v_i$. This hierarchical feature extraction and node update strategy, inspired by the algorithm, is more conducive to the GNN deep cognition of data features. The network architecture is depicted in Figure \ref{Fig.1}.

The first GNN module is used to unroll the equation of the algorithm that calculates $u_i$ and $w_i$. In (\ref{equ.3-10}), the feature of neighbor power $v_j$ and path loss $h_{ij}$ is extracted by $\mathrm{MLP_1}$. To cope with the lack of channel information, we adopt the MAX pooling operation to aggregate the neighborhood information $\alpha_{u_i}$. Feeding $\alpha_{u_i}$ and node feature $v_i,h_{ii}$ into the the combination function $\mathrm{MLP_2}$ in (\ref{equ.3-11}), we can obtain the $u_i$ information of the nodes. Similarly, $w_i$ is calculated by $\mathrm{MLP_3}$ in (\ref{equ.3-12}).

\begin{align}
\centering
 {\alpha_{u_i}}^{(k)}&= \mathrm{MAX}\left\{\mathrm{MLP_1} \left(h_{ij}, {v_j}^{(k-1)}\right )\right\}, j\in \mathcal{N}(i), \label{equ.3-10} \\
 {u_i}^{(k)}&= \mathrm{MLP_2} \left ( h_{ii}, {v_i}^{(k-1)}, {\alpha_{u_j}}^{(k)}\right ) \label{equ.3-11} \\
 {w_i}^{(k)}&= \mathrm{MLP_3}\left ( h_{ii}, {v_i}^{(k-1)}, {u_i}^{(k)}\right ), \label{equ.3-12} 
\end{align}

We utilize the second GNN to unroll equation (\ref{equ.2-9}), which serves as a basis for computing the node power $v_i$. In order to compute the denominator part of (\ref{equ.3-12}), we employ $\mathrm{MLP_4}$ as the aggregation function for the second GNN, which enables us to gather information on neighboring nodes $u_j$, $w_j$, and the edges $h_{ji}$ feature. The neighborhood information $\alpha_{v_i}$ is aggregated by MAX pooling operation. Finally, we use $\mathrm{MLP_5}$ in (\ref{equ.3-13}) as the combining function to update the node power information $v_i$. 
\begin{align}
\centering
 {\alpha_{v_i}}^{(k)}&= \mathrm{MAX}\left\{\mathrm{MLP_4} \left(h_{ji}, {u_j}^{(k)}, {w_j}^{(k)}\right )\right\}, j\in \mathcal{N}(i), \label{equ.3-13} \\
 {v_i}^{(k)}&=\gamma \left(\mathrm{MLP_5} \left( \lambda_i,h_{ii}, {u_i}^{(k)},{w_i}^{(k)},{\alpha_{v_i}}^{(k)}\right)\right) \label{equ.3-14}, 
\end{align}
where $\gamma (x)$ is a sigmoid function to constrain output power, i.e., $\mathcal\gamma (x)=\frac{1}{1+e^{-x}}$. 

\subsection{The Training Approach}

The selection of the loss function for neural networks has a significant impact on the overall performance of the network. In supervised learning, the loss function is computed using the labels $\hat{v}_i$ derived from the WMMSE algorithm. However, in realistic scenarios, communication channel and network topology change quickly, and ground truth labels required for training are difficult to collect within a limited period of time. And it is demonstrated in \cite{r6} that the algorithm's output constrains the upper limit of convergence performance.

The unsupervised loss function in (\ref{equ.14}) uses problem formation to optimize the neural network. Recent research on \cite{r6} \cite{r7} has shown that unsupervised training approaches outperform the WMMSE algorithm. Therefore, for unsupervised training of our model, we adopt the optimization objective as the loss function in (\ref{equ.14}).
\begin{equation}
 \mathcal{L}_{U}(\theta )=-\mathbb E\left(\sum_{i=1}^{N}{ { \lambda_i} \mathrm{log_2}\left(1+\frac{|h_{ii}v_i(\theta )|^2 }{\sum_{i\neq j}^{}|h_{ji}v_j(\theta )|^2 +\sigma^2 }\right)} \right), 
\label{equ.14}
\end{equation}
where $\theta $ is the learnable parameter of the neural network.

\section{Theoretical Assumptions about Validity of Graph Neural Network Unrolling Approach}

In this section, we discuss the validity of deep unrolling techniques for GNNs with respect to network mapping and the size of the model exploration space. Traditional neural networks employ the back-propagation algorithm to determine the steepest gradient descent direction for updating network parameters. The acquired network mapping is optimal for solving optimization problems under a specific data distribution, which can be conceptualized as end-to-end matrix multiplication. However, the learned mapping is highly dependent on the data distribution.

\subsection{Constrained Neural Network Mapping}
As stated in \cite{r15}, a good algorithm alignment means that all algorithm steps of an iterative algorithm are easy to learn. To facilitate learning complex end-to-end mappings, the unrolling approach decomposes the one-iteration process of the algorithm into smaller, simpler subtasks in a hierarchical fashion. Each module learns a portion of the mapping relationship in the optimization algorithm, thereby reducing the overall complexity of model training. For instance, the Newton iterative approach in (\ref{equ.105}) necessitates that an end-to-end neural network learn the entire mapping of $G\left ( x \right)$, containing multiple steps. When $f\left(x\right)$ is complex, learning the end-to-end mapping is challenging. By dividing the mapping into distinct $g_i\left ( x \right )$ modules in (\ref{equ.107}), the unrolling approach can learn these mappings, including crucial parameters such as iteration steps during the iterative process. This technique effectively learns the algorithmic mapping and, through multi-level mapping, enhances the expressive network capability compared to one-step mapping.
\begin{align}
\centering
&\;\;\;\;\;\;\;\;\;\;\;x_{n+1} = x_{n}-\frac{f\left(x_{n}\right)}{f^{\prime}\left(x_{n}\right)} \label{equ.105}\\ 
&\;\;\;\;\;\;\;\;\;\;\;\Leftrightarrow G\left ( x \right ):x_{n+1}\to x_{n}\label{equ.106}
\\
&\;\;\;\;\;\;\;\;\;\;\;\Leftrightarrow \begin{cases}
g_1\left ( x \right ):x_{n}\to f\left(x_{n}\right)
\\ g_2\left ( x \right ):x_{n}\to f^{\prime}\left(x_{n}\right)
\\g_3\left ( x \right ):\left ( f^{\prime}\left(x_{n}\right),f\left(x_{n}\right),x_{n} \right ) \to x_{n+1}
\end{cases}
\label{equ.107}
\end{align}

\subsection{Reduce the Model Exploration Space}

Furthermore, the input-output relationship between algorithm-designed modules constrains the mapping direction, and algorithm-based feature extraction aids in reducing the network exploration space. Building on the conclusions in \cite{r14} that deeper GNNs increase feature correlations, we examine the impact of input inter-feature correlations on network exploration. Employing the widely used Pearson correlation coefficient measure, \cite{r14} introduces the metric $Corr(X)$ to quantify the correlation between all feature dimension pairs. 
\begin{equation}
\small
{Corr(\mathbf{X})=\frac{1}{d(d-1)}\sum_{i\neq j}^{}|\varphi \left (\mathbf{X}_{\left (:, i\right )}, \mathbf{X}_{\left (:, j\right )}\right )|, 
\; i, j \in [1,2,\dots,d],} 
\label{equ.13}
\end{equation}
where $\mathbf{X} \in \mathbb{C}^{N\times d}$ represents the node feature matrix with $d$ being the number of features, $\mathbf{X}_{\left (:, i\right )}$ indicates the $i$-th column feature vector of $\mathbf{X}$, and $\varphi (x, y)$ is a popular Pearson correlation coefficient for assessing the correlation between learned dimensions in deep GNNs.

We propose that the exploration space of a neural network can be viewed as a high-dimensional space, expanded orthogonally by independent input variables. Considering two one-dimensional features on a graph node, it is observed that when the two input features are independent, the exploration space can be characterized as a high-dimensional space $\mathbb{R}^{2}$, expanded orthogonally by two one-dimensional variables. In contrast, if the two features are identical, the exploration space is reduced from a high-dimensional, orthogonally expanded $\mathbb{R}^{2}$ space to a one-dimensional linear expanded $\mathbb{R}^{1}$ space. This reduction in the exploration space size is manifested by an increase in the correlation coefficient $Corr(X)$ between input features, with $Corr(X)$ varying from 0 to 1. Therefore, the relationship between exploration space and input feature correlation can be expressed as $\mathbb{R}^{{d}^{1-Corr(X)}}$.

Based on an iterative algorithm, the unrolling approach decomposes end-to-end optimization in a single iteration into multiple sub-problems that are iteratively processed. Through repeated feature extraction, this approach increases the correlation between node features in GNNs. For example, in the WMMSE algorithm, the $k$-th power $v_{i}^{(k)}$ is generated by channel features $h$ plus the power $v_{i}^{(k-1)}$ from the previous iteration, which creates a certain correlation between the input features of the GNN. Moreover, in the deep unrolling architecture, two message passing operations and multiple channel information pass to increase the correlation between power $v_i$ and channel characteristic $h$. This approach is more effective in feature extraction with higher correlation of features after multi-layer networks and more accurate exploration of optimization objectives, especially when faced with changing data distributions.

\section{Numerical Experiments}
This section is dedicated to the conduction of comprehensive numerical tests to affirm the effectiveness and generalization of the presented knowledge-driven network architecture. Our experimental setup consists of a user count set to $N=10$, a noise variance of $\sigma^2=10$dB, and an interference channel following a Rayleigh distribution. We derive the channel coefficients $h_{ij}$ from the complex normal distribution $\mathcal{CN}(0, 1)$. In terms of the neural network training parameters, we utilize the Adam optimizer with a set learning rate of $0.001$ and designate the batch size to be $64$ samples. We compare UWGNN with established benchmarks and cutting-edge approaches.
\begin{enumerate}[]
\item \textbf{WMMSE}{
\cite{r2}: This is a classical iterative optimization algorithm for weighted sum rate maximization in interference channels. We run WMMSE for 100 iterations with $p_\mathrm{max}$ as the initial power setting. The results obtained from this process served as our benchmark measurements.
}
\item \textbf{WCGCN}{
\cite{r9}: This is an unsupervised message passing GNN that uses two MLP networks to aggregate neighbor information and update its power information, and obtain a performance much better than the WMMSE algorithm.
}
\item \textbf{UWMMSE}{
\cite{r13}: UWMMSE proposed a deep unrolling architecture based on GNN to learn the iterative step-size $a^{(k)}$ and the weight translation parameter $b^{(k)}$, reduced the times of WMMSE iterations, and attained the performance comparable with well-established benchmarks.
}
\item \textbf{MLP}{
\cite{r4}: MLP uses the WMMSE output as a training label to supervise and learn a function mapping between the channel state information and the corresponding resource allocation.
}
\end{enumerate}

We set UWGNN, WCGCN, and UWMMSE as three-layer networks, corresponding to three iterations, and use the same random seed to generate $10^4$ sets of channel training samples to train three networks, respectively. Learning the iterative process with an MLP can be challenging and may require more training samples. Therefore, to effectively train the model, we utilized a ten-fold increase in training data in this study.

\subsection{Selection of UWGNN Hyperparameters}

\begin{figure}[htbp]
\centering 
\subfigure[]{
\label{Fig9. sub. 1}
\includegraphics[width=0.35\textwidth]{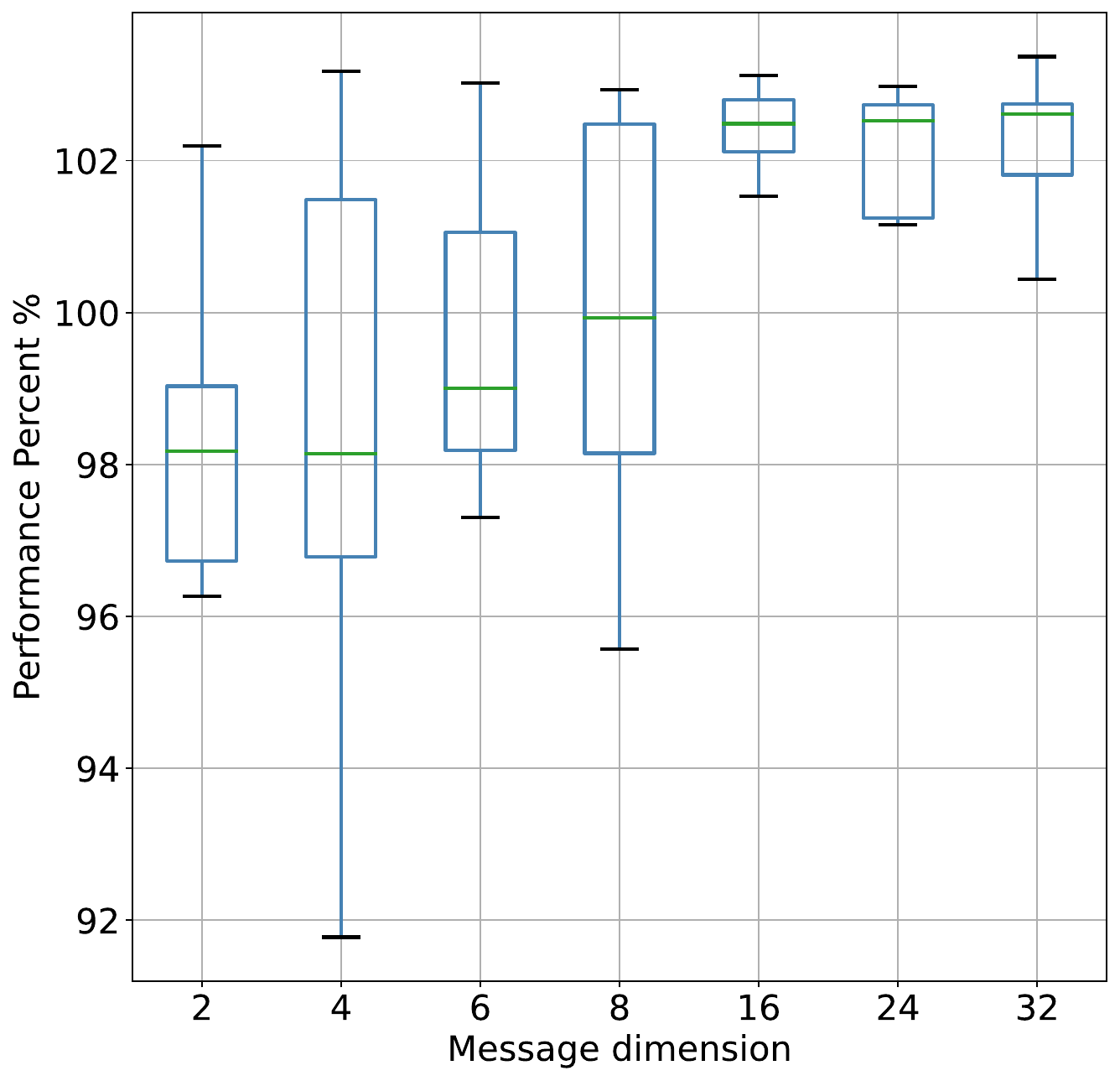}}
\subfigure[]{
\label{Fig9. sub. 2}
\includegraphics[width=0.35\textwidth]{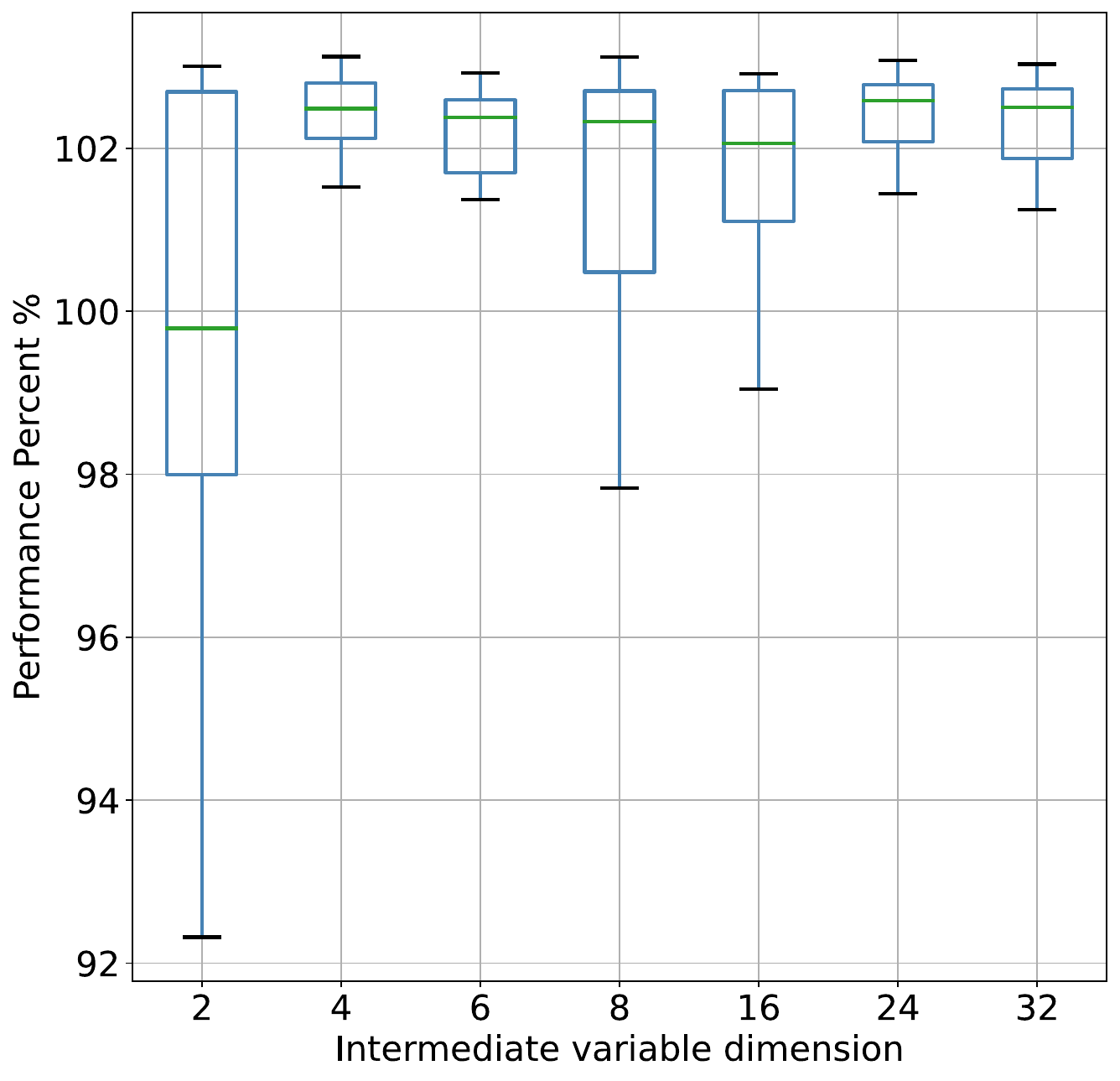}}

\caption{Impact of variation in width of GNN on performance}
\label{Fig.9}
\end{figure}
In this section, our investigation centers on how model performance is influenced by factors such as the intermediate message-passing dimension, which is the output dimension of $\mathrm{MLP_{1,4}}$, the variable dimension, which corresponds to the output dimension of $\mathrm{MLP_{2,3}}$. We used 20 random seed variables in our experiments and chose the WMMSE algorithm output as a benchmark. As illustrated in Fig. \ref{Fig9. sub. 1}, the effect of the message passing size on model performance was demonstrated. We observed that when the dimension of the aggregation message was too small, UWGNN performance declined. When the dimension increased to 16, UWGNN performance exceeded the WMMSE baseline. Further increases did not impact the model performance significantly. We hypothesized that the message passing dimension influences the ability of the node to extract information from neighboring nodes, as evidenced by (\ref{equ.3-14}), and concluded that the dimension of message variables must be at least equal to the sum of the dimensions of $u_j$, $w_j$, and $h_{ij}$. Reducing message passing dimension will lead to feature information compression of neighboring nodes that decrease network performance. On the other hand, excessively wide message passing dimension only marginally enhanced model performance but generated computational complexity redundancy.


For the intermediate variable dimensions, as presented in Fig. \ref{Fig9. sub. 2}, the impact on the performance of the network model  is minimal, with a slight decrease in performance observed only for intermediate variable dimensions equal to two. The intermediate variables designed in our study, based on one-dimensional messages $u_i$ and $w_i$ in (\ref{equ.2-7})-(\ref{equ.2-9}). Increasing the dimension of these variables has little impact on one-dimensional feature extraction. Thus, the size of the message passing dimension in Fig. \ref{Fig.9} does not significantly affect network performance. However, in distributed GNNs, communication resources are utilized during the message passing process. An excessively wide intermediate variable dimension requires a larger message passing dimension, which can increase transmission latency and path loss, thereby affecting the accuracy and delay of GNN inference. As a result, we aim to strike a balance between network feature extraction and optimal network performance by ensuring that the feature dimension of $u_i, w_i$ is comparable to the input dimension of $h_i, v_i$. 



As stated above, we establish the network unit size of $\mathrm{MLP_1-MLP_5}$ in (\ref{equ.3-10})-(\ref{equ.3-14}) to \{5, 8, 16\}, \{19, 8, 4\}, \{7, 8, 4\}, \{10, 8, 16\}, and \{27, 8, 1\}. UWGNN is constructed with three unrolled WMMSE layers, each of which shares parameters. In every layer, two GNNs are incorporated as learning components, utilized for the emulation of (\ref{equ.2-7})-(\ref{equ.2-9}).

\subsection{Sum Rate Performance}

\begin{table}[htbp]
\centering
\caption{Sum Rate Performance}
\label{table. 1}
\scalebox{0.8}{
\begin{tabular}{|c|c|c|c|c|l|}
\hline
\begin{tabular}[c]{@{}c@{}}User \\ numbers\end{tabular} &
 UWGNN &
 WCGCN &
 UWMMSE &
 MLP &
 \multicolumn{1}{c|}{\begin{tabular}[c]{@{}c@{}}WMMSE \\ 100 times \end{tabular}} \\ \hline
10 & 102.817\% & 102.313\% & 105.302\% & 94.794\% & 105.801\% \\ \hline
30 & 102.996\% & 102.094\% & 92.168\% & 71.476\% & 102.730\% \\ \hline
50 & 102.256\% & 101.743\% & 90.497\% & 60.485\% & 102.493\% \\ \hline
\end{tabular}}
\end{table}

We compare the sum rate performance obtained by UWGNN with other approaches, as shown in Table \ref{table. 1}. To determine the upper bound, we ran the WMMSE algorithm 100 times for random power initialization and selected the best performance. As shown by the results, both UWGNN and WCGCN are in close proximity to the benchmark performance when dealing with relatively smaller problem scales. As the user size increases, MLP is not an optimal choice because it requires a large number of training samples. This observation indicates that MLP is not well-suited for learning an iterative process. In contrast, UWGNN and WCGCN demonstrate attractive properties as their performance remains stable and close to the performance ceiling of WMMSE, even as the user size increases. It is worth noting that UWMMSE requires the WMMSE algorithm to obtain the output power, which significantly impacts its performance as the computational complexity increases. Our findings suggest that message passing GNNs are preferable and more effective than other approaches for the iterative optimization problems.

\subsection{Convergence Speed Comparison}
\begin{figure}[http]
\centering
\includegraphics[width=0.40\textwidth]{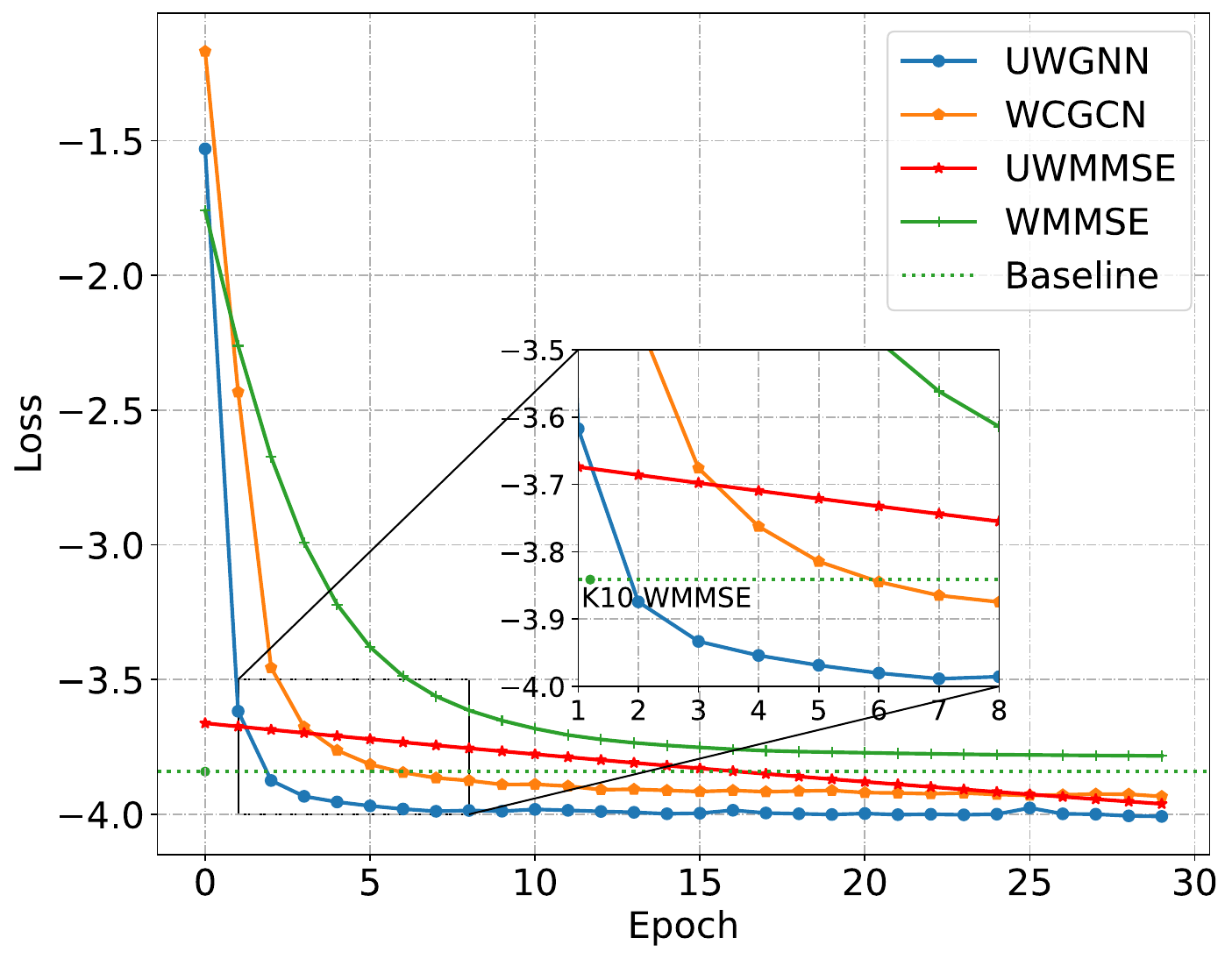}
\caption{Network convergence speed comparison}
\label{Fig.2}
\end{figure}

\begin{figure*}[htbp]
 \centering 
 \subfigure[Variance alteration in Rayleigh channel]{
 \label{Fig4. sub. 1}
 \includegraphics[width=0.28\textwidth]{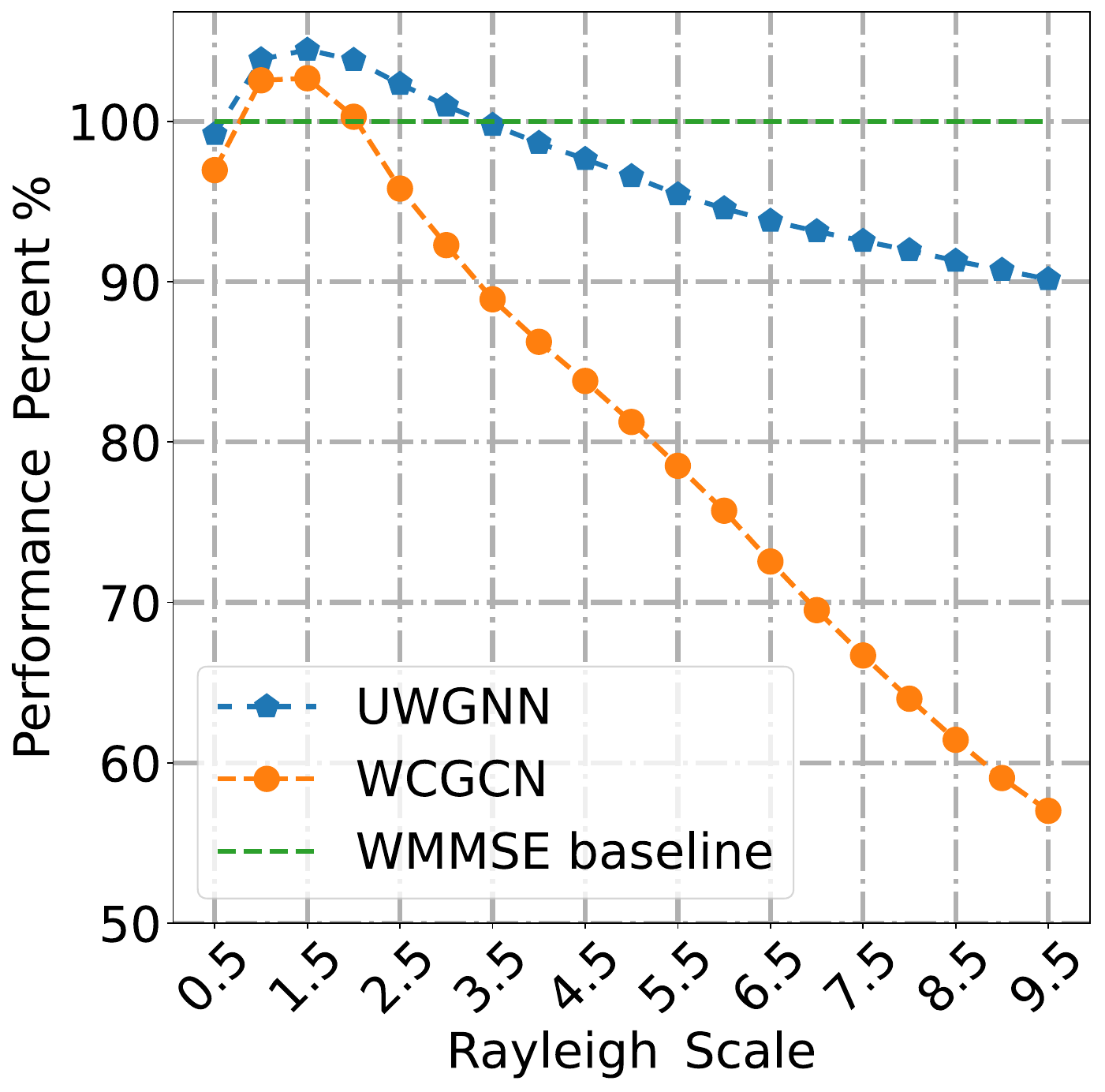}
 }
 \subfigure[Transition from Rayleigh to Rician channel and variance shift]{
 \label{Fig4. sub. 2}
 \includegraphics[width=0.28\textwidth]{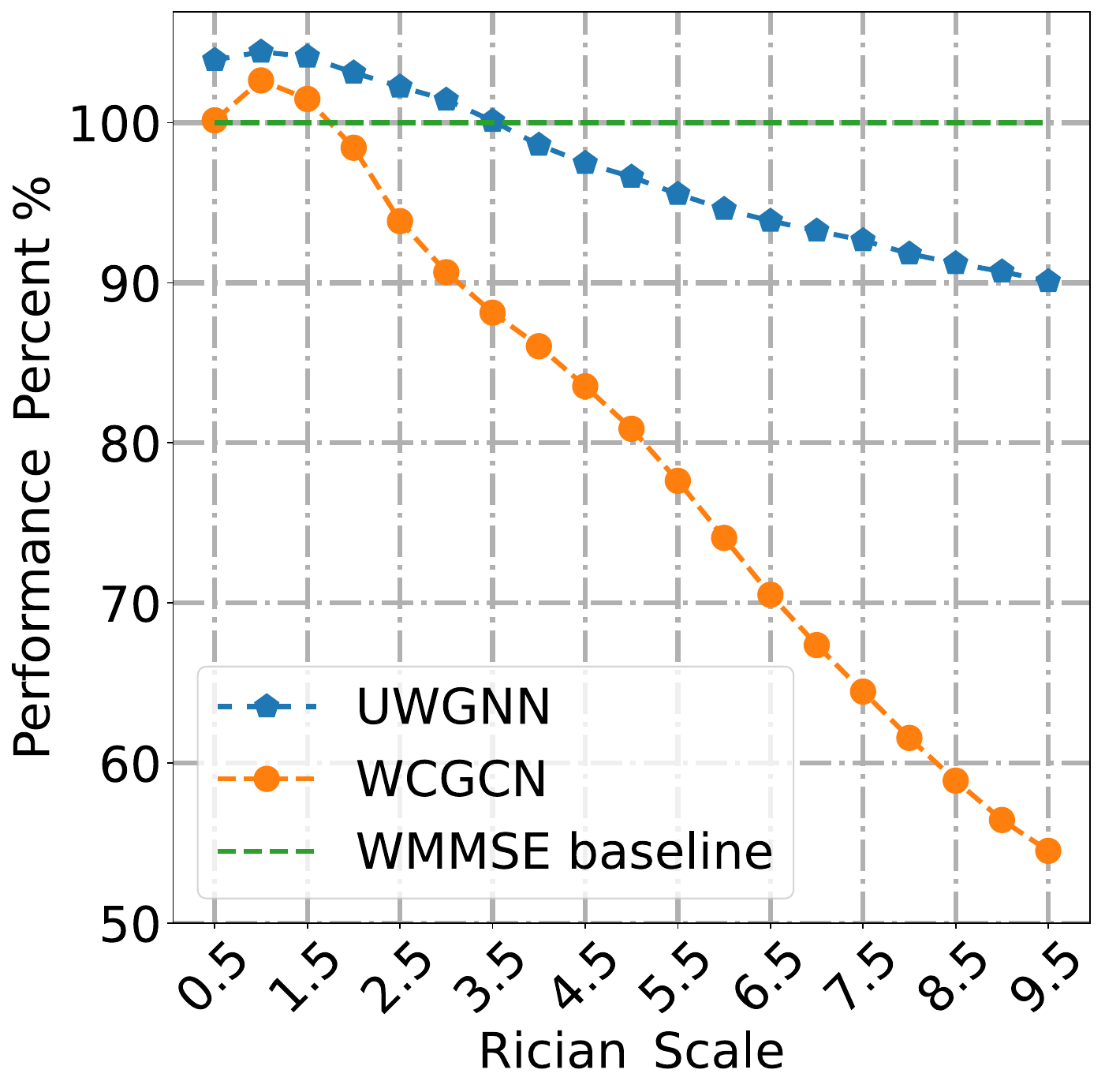}
 }
 \subfigure[Adjustment the strength of LOS component]{
 \label{Fig4. sub. 3}
 \includegraphics[width=0.28\textwidth]{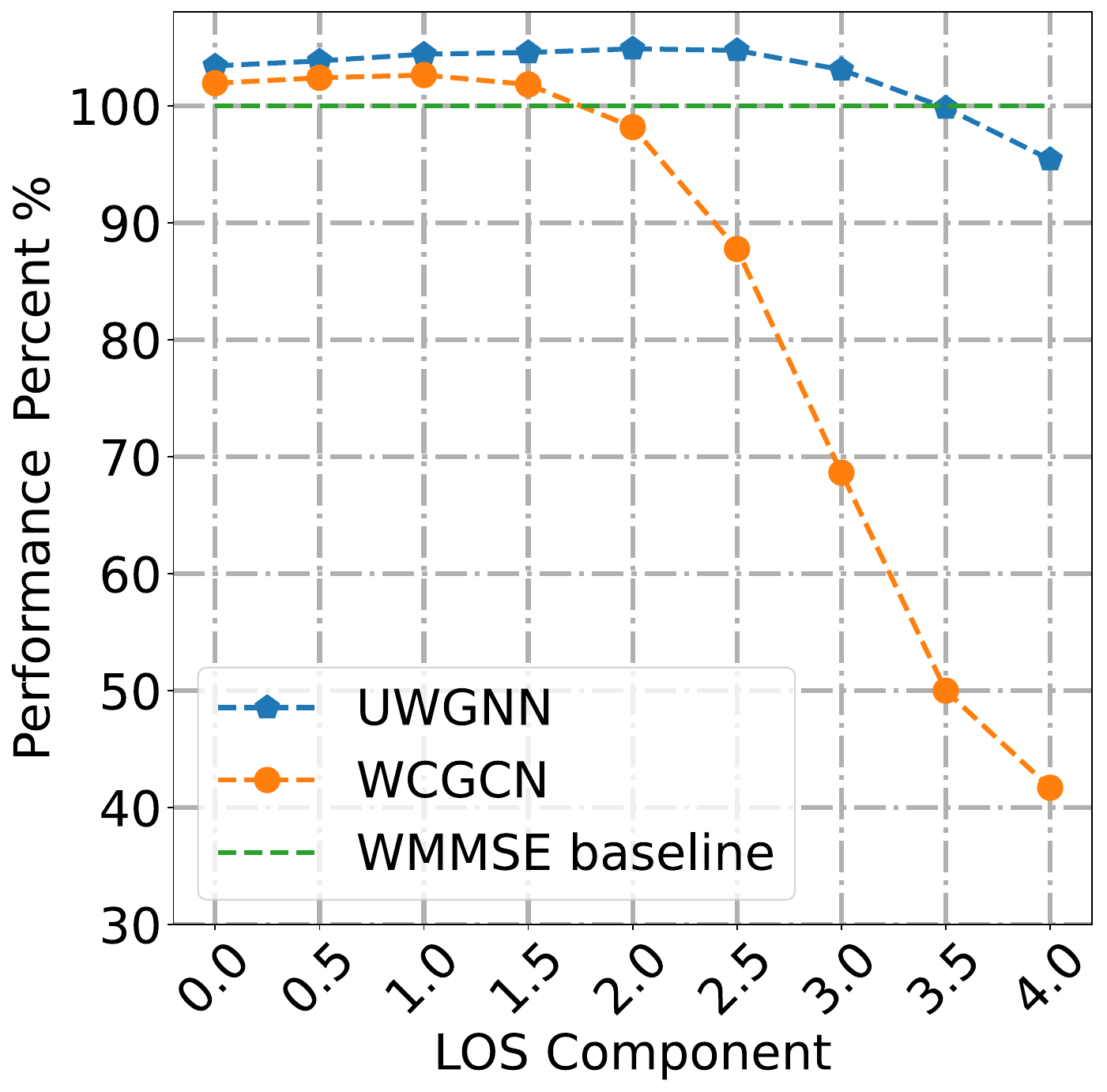}
 }
 \caption{Channel distribution generalization comparison}
 \label{Fig.4}
\end{figure*}
We compare the network convergence rate in Fig. \ref{Fig.2}, and our network, which converges in only $\frac{1}{3}$ of the epochs to achieve convergence performance compared to that of WCGCN. UWMMSE learns the step size parameter and uses the WMMSE algorithm to iteratively obtain the output. Therefore, the initial performance of UWMMSE is better, but it presents a slower linear convergence speed. Although the deep unrolling operation increases the computational complexity of the network from $O(L(|E|+|N|))$ to $O(L(|E|+|N|^2 )(|E|+|N|))$, our network architecture is narrower in width and deeper in depth; thus the network parameters of UWGNN has a smaller parameter count than WCGCN. We employ the thop library in Python for comparing the computational load and the network's parameter count, as specified in Table \ref{table. 2}. The proposed network converges faster because we have more iterations of channel state information and node power information, resulting in a smaller exploration space, and the same feature extraction approach as the WMMSE algorithm. 
\begin{table}[htbp]
\centering
\caption{Network Computation and Parameters}
\label{table. 2}
\scalebox{0.8}{
\begin{tabular}{|c|c|c|}
\hline
Network & MACs {(G)} & Params {(K)} \\ \hline
UWGNN 3layers & 1.181 & 1.906 \\ \hline
WCGCN 3layers & 1.797 & 2.466 \\ \hline
\end{tabular}}
\end{table}
\begin{figure}[htbp]
\centering
\includegraphics[width=0.4\textwidth]{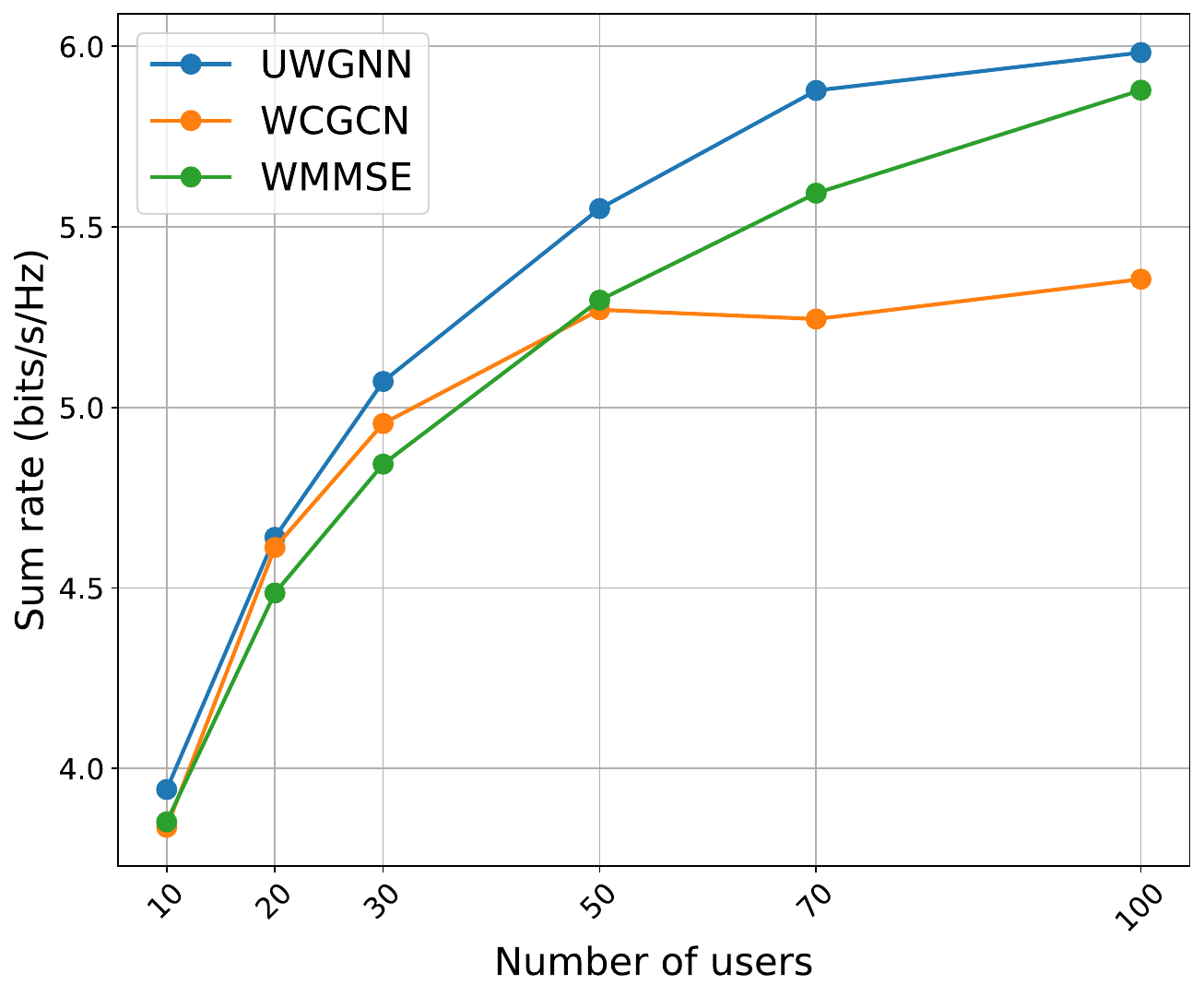}
\caption{Scalability comparison}
\label{Fig.3}
\end{figure}
\vspace{-0.66cm}
\subsection{Scalability Comparison}
In order to assess network scalability, both UWGNN and WCGCN were independently trained over 30 epochs until convergence was achieved in a scenario with 20 users. Subsequently, these trained networks were transferred to new scenarios featuring varying numbers of users, without the need for additional training. As shown in Fig. \ref{Fig.3}, both GNN networks have good convergence performance for a smaller number of 10 user scenario. However, when the number of users increases to 50, the WCGCN network can no longer exceed the performance of the WMMSE baseline, while our network can still exceed the baseline performance. As the number of users reaches 100 and the user connection density intensifies, the WCGCN can no longer achieve the baseline performance, while our network architecture still achieves the baseline performance.

After our experiments, we found that the scalability of GNN comes from the smoothing operation of the max pooling layer on the features of neighboring nodes. When the user dimension changes, the pooling layer uses $\mathrm{SUM(.)}$, $\mathrm{MAX(.)}$, or $\mathrm{MEAN(.)}$ functions to extract the feature information of neighbor nodes and edges. Experimentally, it is found that the MAX function works best for the problem in this paper. Moreover, our network architecture undergoes two rounds of MAX pooling because of two information aggregation operations; thus, it is better suited for scaling to diverse scenarios with varying user number densities. 
\begin{figure*}[htbp]
\centering 
\subfigure[Dense to sparse unweighted ]{
\label{Fig6. sub. 1}
\includegraphics[width=0.23\textwidth]{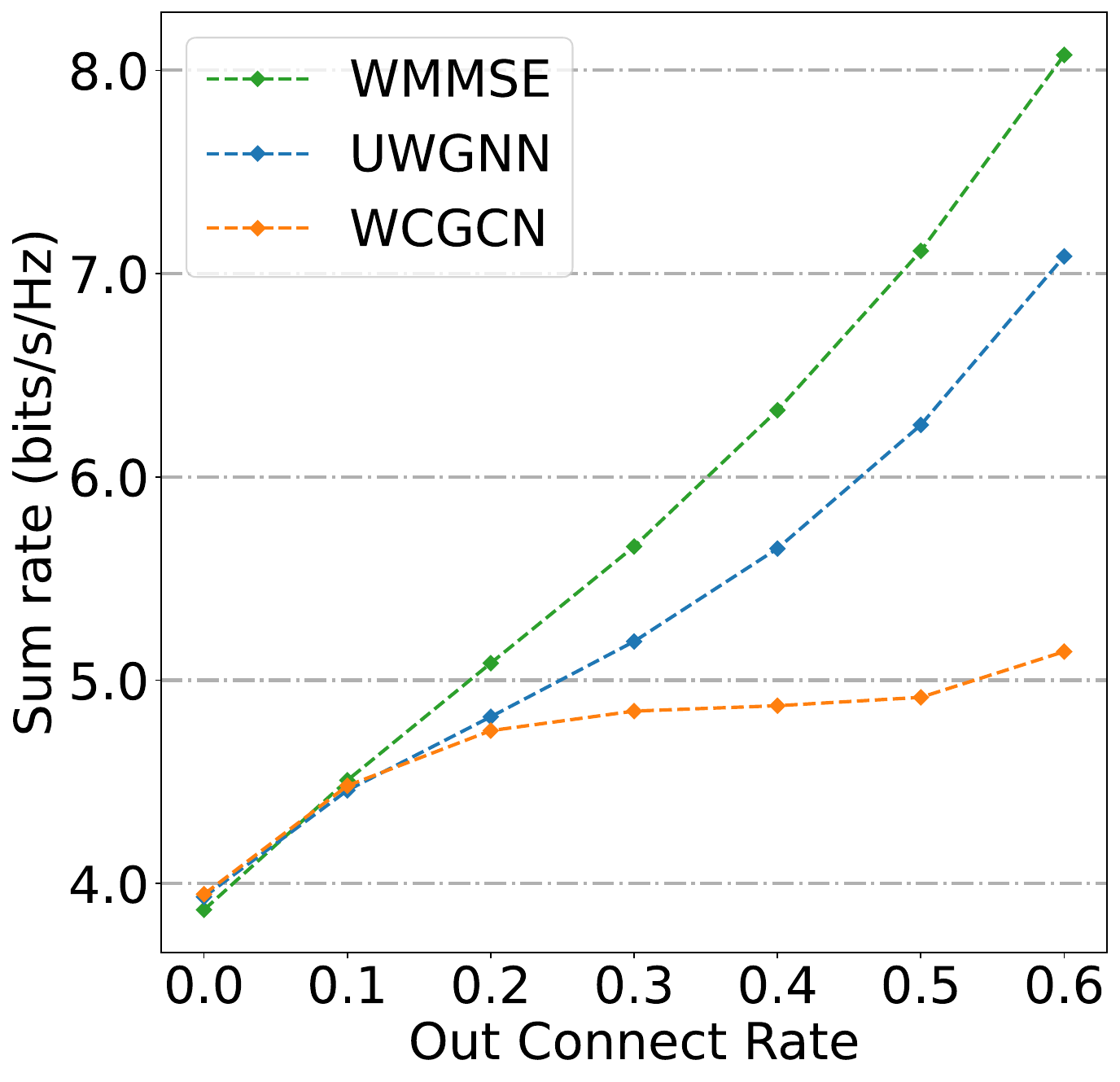}}
\subfigure[Dense to sparse weighted]{
\label{Fig6. sub. 2}
\includegraphics[width=0.23\textwidth]{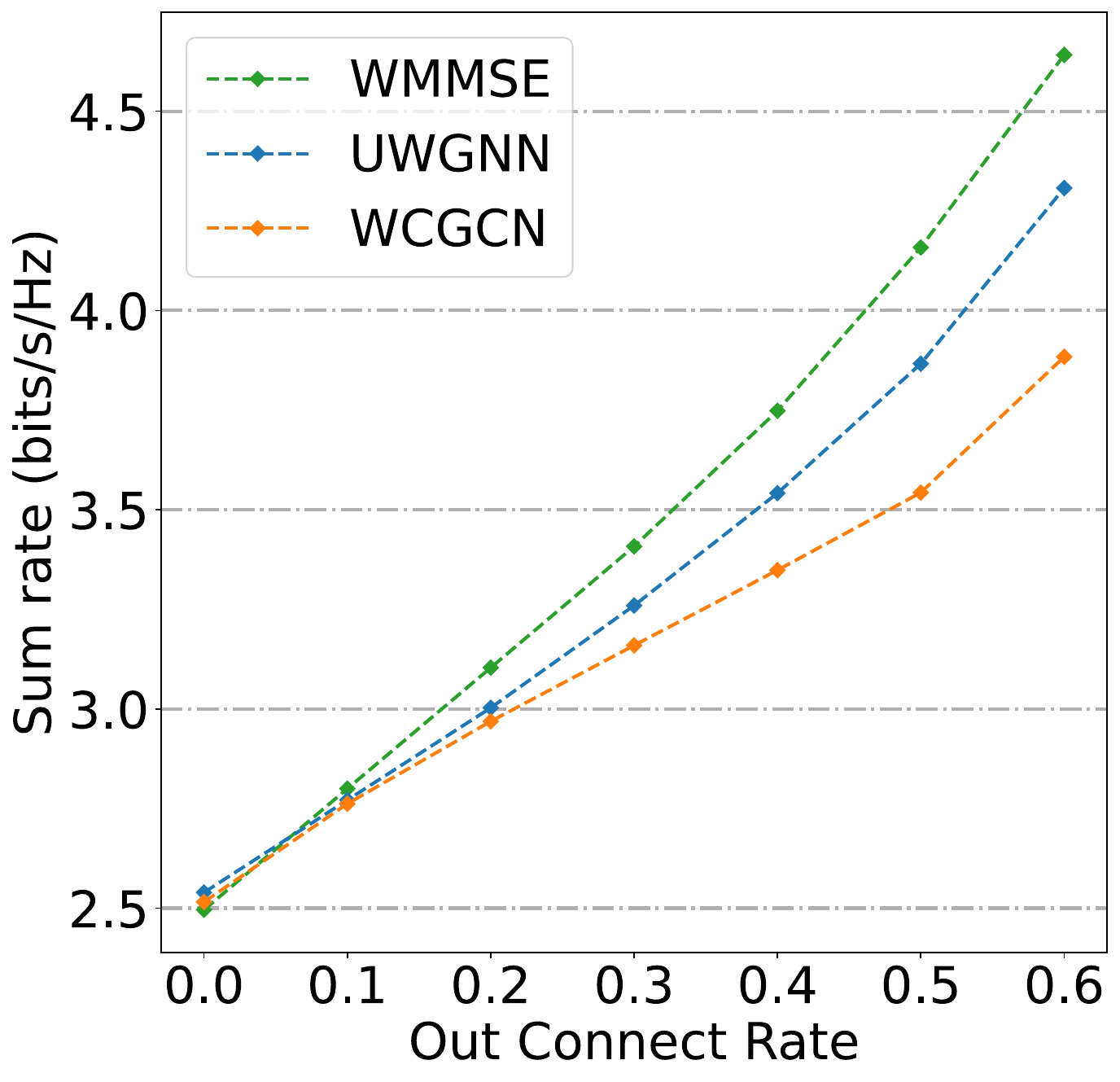}}
\subfigure[Sparse to dense unweighted]{
\label{Fig6. sub. 3}
\includegraphics[width=0.23\textwidth]{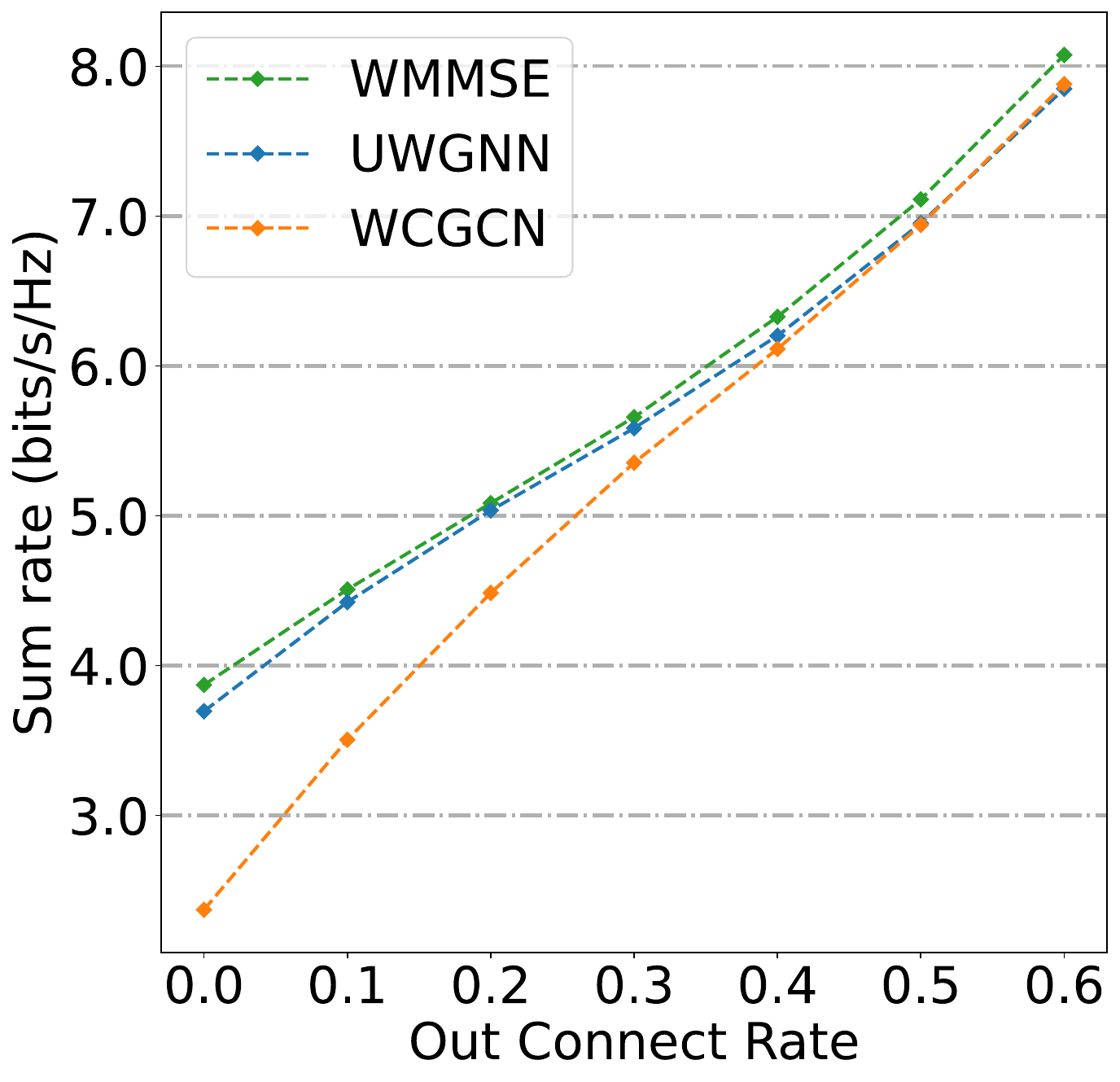}}
\subfigure[Sparse to dense weighted]{
\label{Fig6. sub. 4}
\includegraphics[width=0.23\textwidth]{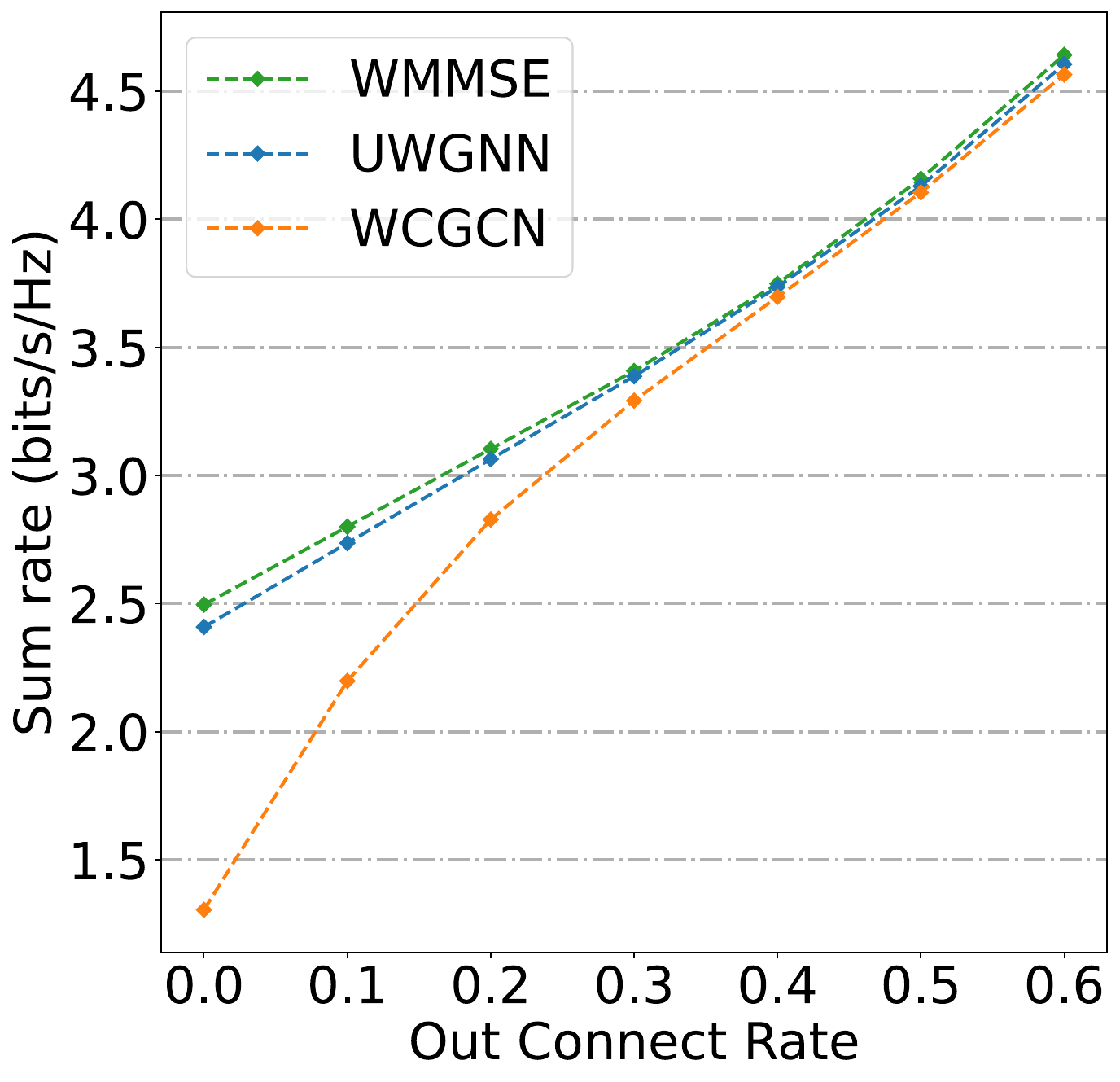}}
\caption{Performance comparison of changing communication network topology}
\label{Fig.6}
\end{figure*}
\begin{figure*}[]
\centering
 \includegraphics[width=0.9\textwidth]{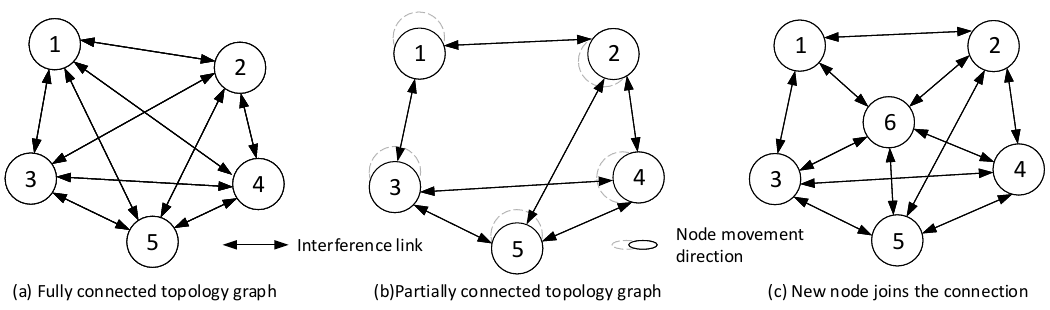}
 \caption{Change the communication network topology. With the movement of communication nodes, the interference between certain nodes in the fully connected communication topology will decrease, and the communication network topology will change. The addition of new nodes will also change the communication topology.}
 \label{Fig.5}
\end{figure*}

\subsection{Channel Distribution Generalization}

The generalization performance of the network is tested using datasets with varying channel distributions. As the user moves, the scattered channel may convert into a direct channel, leading to a potential shift in data distribution from Rayleigh to Rician distributions in the communication scenario. To ascertain the model's generalization ability, we adjusted the sample distribution of the channel in the test set. The initial training set consisted of Rayleigh channels with a mean of 0 and a variance of 1. As shown in Fig. \ref{Fig4. sub. 1}, we modified the variance of the Rayleigh channel within the test set. During minor variance changes, UWGNN and WCGCN both exhibited generalization capabilities. Nonetheless, as the variance gap widened, WCGCN's performance deteriorated noticeably, signifying its limitations in adapting to data distribution shifts. With the integration of a knowledge-driven network grounded on the WMMSE algorithm, our network architecture demonstrated superior generalization compared to WCGCN, maintaining robust adaptability amidst diverse data distributions. Remarkably, our model sustained roughly 90\% of its performance even amidst substantial variance alterations. In Fig. \ref{Fig4. sub. 2}, we introduced the line of sight (LOS) component to the Rayleigh channel by increasing the channel mean to 1, thus transforming its distribution into a Rician distribution. Following this, we altered the variance of the Rician distribution. The experimental results underscored the continued robust generalization performance of our network. Fig. \ref{Fig4. sub. 3} illustrates the impact of modifying the strength of the LOS component within the Rician channel. Our model maintained satisfactory performance under varied direct path strengths, facilitating a seamless transition between scattered and direct channels - a critical feature ensuring the model's generalization ability in a mobile environment.

The conducted experiments affirm that the integration of a knowledge-driven model significantly bolsters generalization across disparate sample distributions. Facing diverse data distributions, our model demonstrates minimal performance degradation, obviating the need for additional retraining.

\subsection{Communication Topology Generalization}

As shown in Fig. \ref{Fig.5}, the topology of a communication network changes with the movement of users and the addition or removal of nodes, which affects the size of the node degree in the graph and thus impacts the performance of GNNs. To evaluate network generalization in adapting to communication topology changes, we generated a connection weight matrix as depicted in \ref{equ.21}. The edges with weights lower than the probability of losing connection $\eta_{lc}$ were removed to simulate sparse connections, allowing us to convert between fully and sparsely connected interference graph datasets. 
\begin{equation}
\centering
C_{(j,i,:)} = \left\{
 \begin{matrix}
 \mathbf{0},\qquad{\mathrm{c_{ji}<\eta_{lc}}} 
 \\ 
 \:\:\mathbf{1},\qquad{\mathrm{otherwise}}
 \end{matrix}\right.
\label{equ.21}
\end{equation}
\begin{equation}
\centering
\hat{A}_{(j,i,:)} =A_{(j,i,:)}\odot C_{(j,i,:)}
\end{equation}
where $c_{ji}$ followed standard Gaussian distribution. $\hat{A}_{(j,i,:)}$ is a new adjacency matrix of communication graph.

We experimented with two training and testing directions: from dense to sparse, and from sparse to dense. First, we trained the network on a fully connected graph data set with 10 pairs of users, migrating the test to sparsely connected graph data. As shown in Fig. \ref{Fig6. sub. 1}, our network degrades in performance as the communication topology becomes more sparse, but still maintains a generalization performance of over 85\%. In contrast, the performance of WCGCN degrades more significantly as the topology of the communication network changes. Compared with the unweighted scenario, the performance of the neural network trained in the weighted summation rate scenario is more generalized in Fig. \ref{Fig6. sub. 2}. This is because the weighted and rate scenarios are more complex and the network learns more data about distribution during training. So when the network topology is changed, it is more adaptable. Second, we train the neural network on a training sample with $\eta_{lc}$ of 0.6 and gradually decrease the $\eta_{lc}$ value on the test data, making the sparse connected graph into a fully connected graph. The results are shown in Figs. \ref{Fig6. sub. 3} and \ref{Fig6. sub. 4}. With increasing communication topology density, our network performance still closely follows the WMMSE algorithm performance, but the traditional end-to-end GNN performance decreases as increasing interference density.

\begin{figure*}[httb]
\centering 
\subfigure[Speed 50 m/s]{
\label{Fig10. sub. 1}
\includegraphics[width=0.28\textwidth]{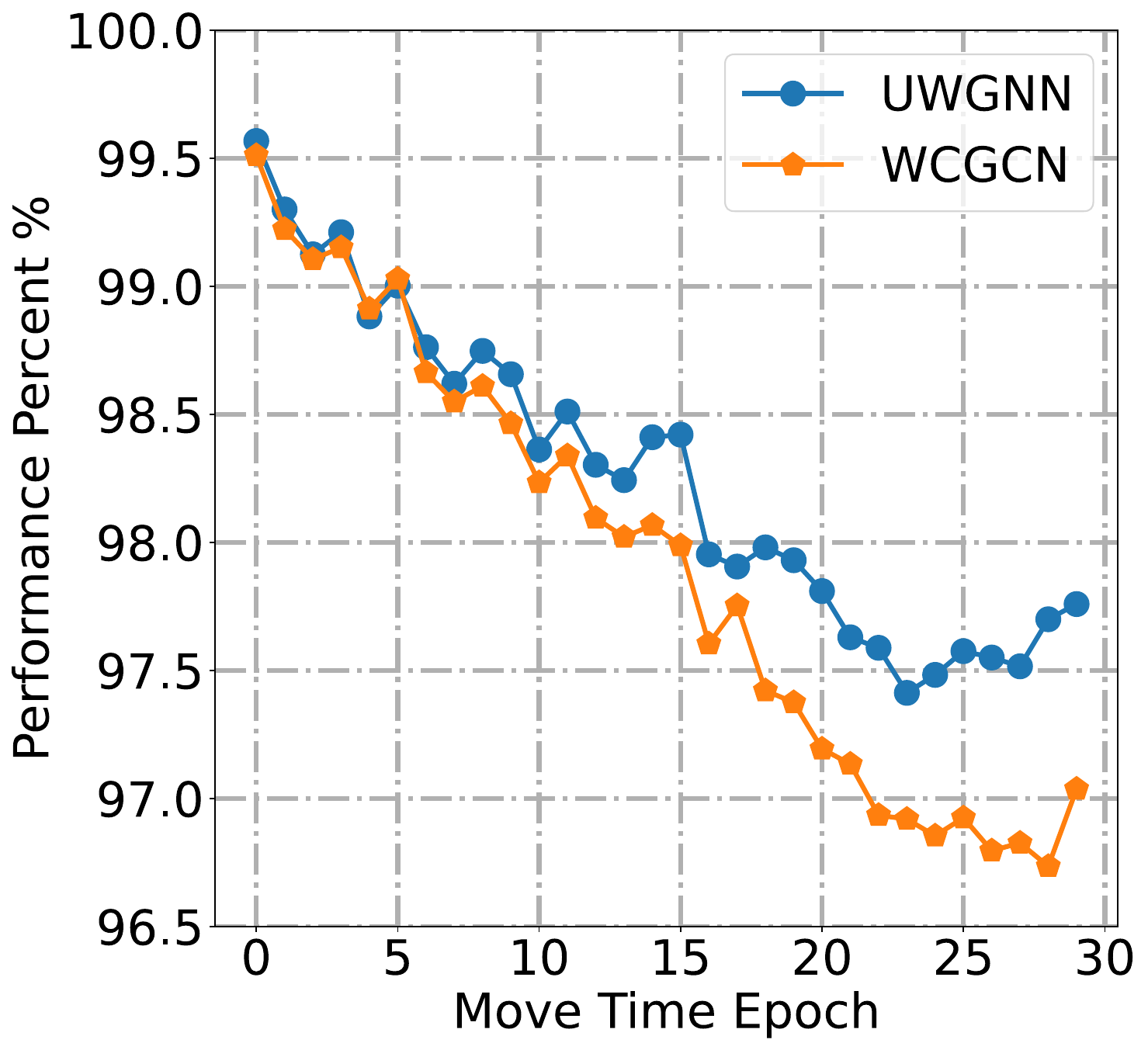}}
\subfigure[Speed 100 m/s]{
\label{Fig10. sub. 2}
\includegraphics[width=0.28\textwidth]{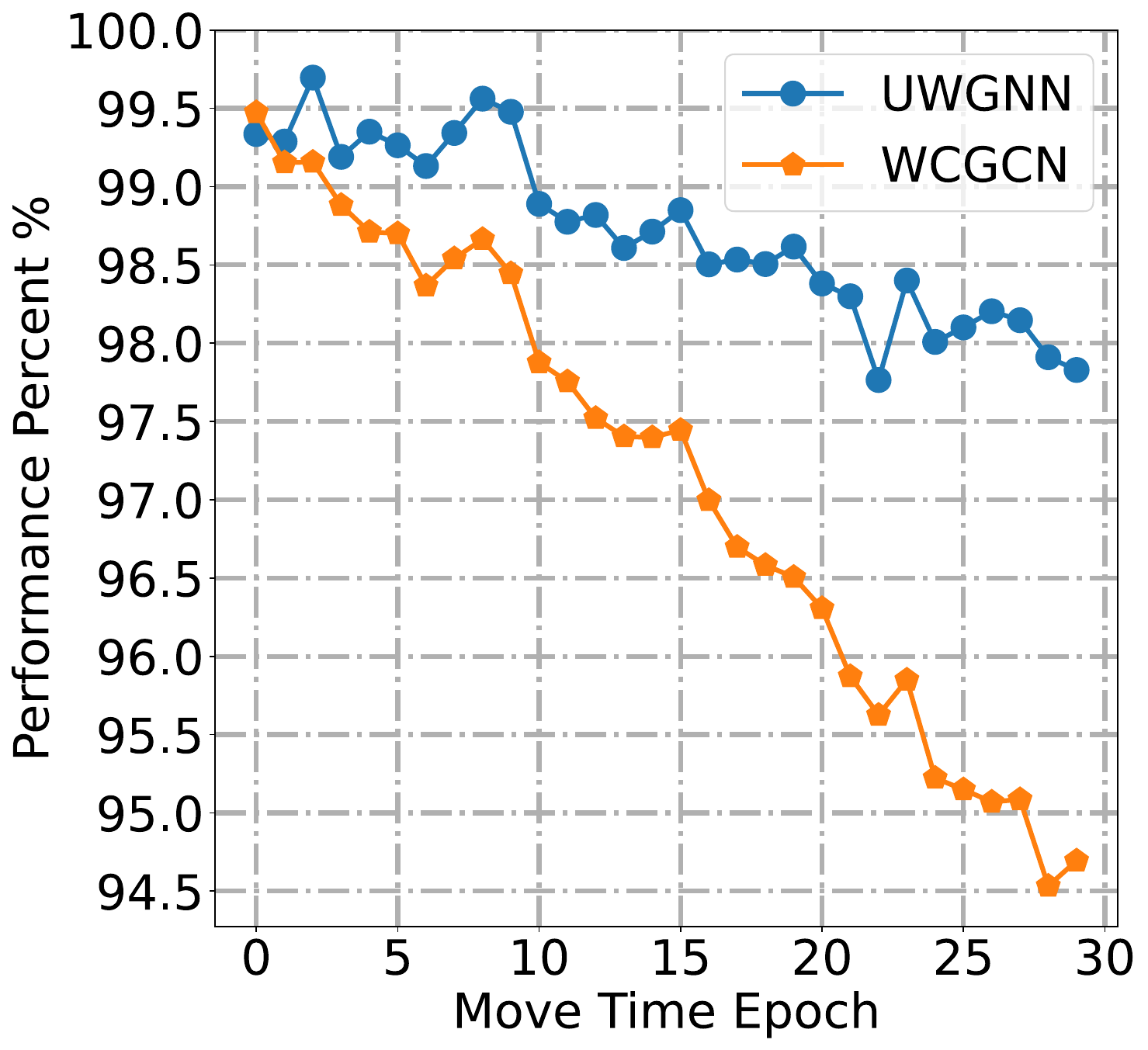}}
\subfigure[Speed 200 m/s]{
\label{Fig10. sub. 3}
\includegraphics[width=0.28\textwidth]{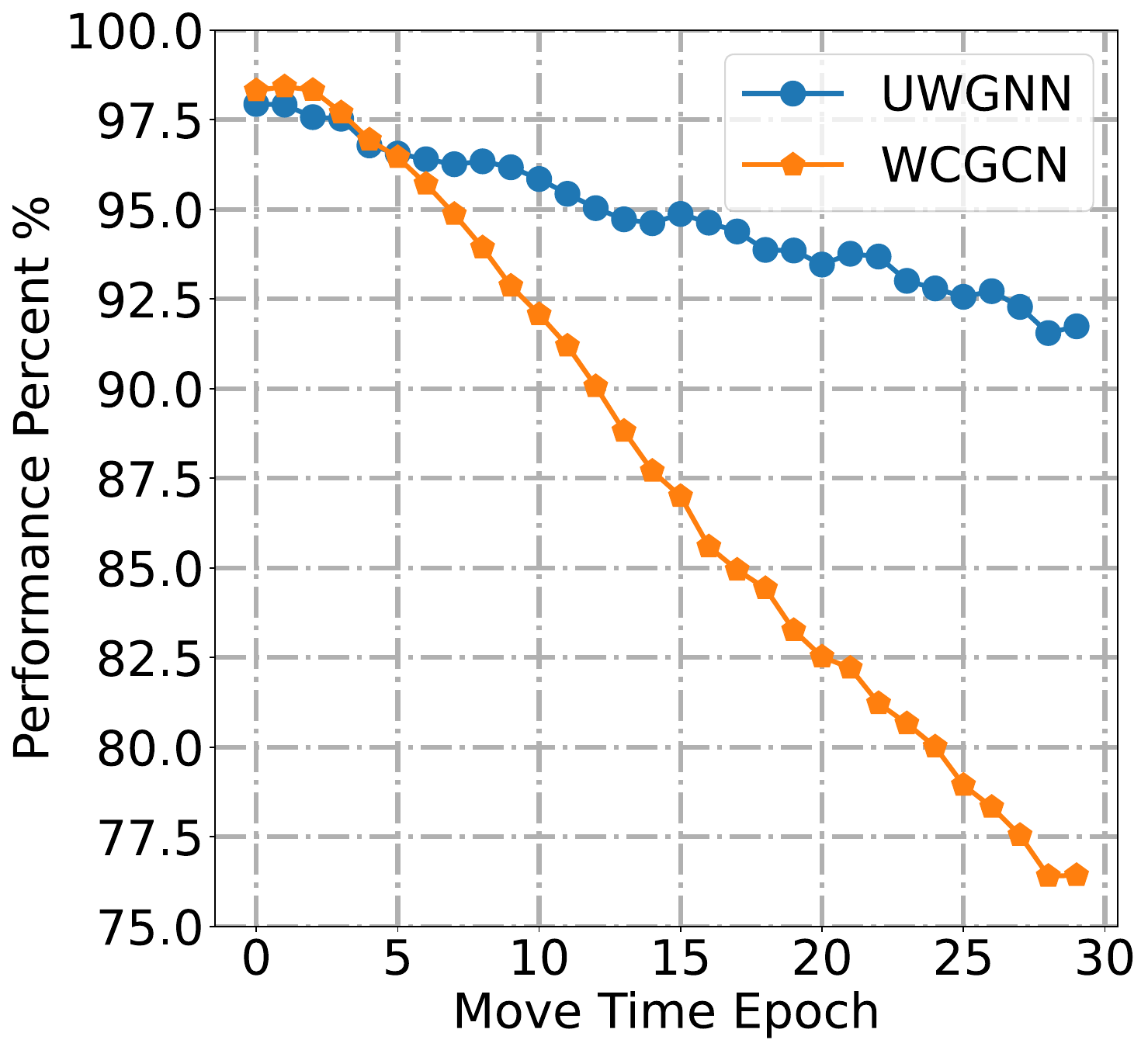}}
\caption{Performance comparison of changing communication network topology}
\label{Fig.10}
\end{figure*}

\subsection{Mobile Generalizability Performance}
The previous experiment only changed the link structure of the communication topology, however, as wireless devices usually move, the communication channel gain changes as well. In this experiment, we distribute N-transmitters uniformly in the space of [$1000\mathrm{m} \times 1000\mathrm{m}$], and the corresponding receivers are distributed around the transmitters with distances obeying uniform distribution $\mathbf{U} (30,90)$. Then we let each receiver device move randomly with speed $S$. The change of position of the receiver obeys a two-dimensional Gaussian distribution $N(0, 0, S, S, 0)$. The distance matrix $\mathbf{D}$ can be obtained by calculating the inter-device distance. we multiply the distance matrix $\mathbf{D}(t)$. The channel gain matrix $\mathbf{H}(t)$ is adjusted proportionally based on distance to reflect the changes in device movement, resulting in the updated channel gain matrix $\mathbf{H}(t+1)$. If the inter-device distance is greater than $1000$ m, we consider that the channel gain is small and set $h_{ij}$ to 0 to simulate the process of moving the user out of the device coverage. In this way, the test sample distribution will gradually move away from the training sample distribution with the increase of move time. 

In Fig. \ref{Fig.10}, we test how the different user movement speeds affect the network generalization. When the device moves at low speed, the communication topology and channel gain change slowly, and the distribution of training and test data do not differ much. So, both GNNs can maintain more than 95\% performance. However, when the speed of device movement gradually increases, and the test data changes more drastically, the end-to-end GNN has difficulty in adapting to the new distribution and the performance degrades. Due to the incorporation of algorithmic knowledge, our proposed GNN excels in its generalization capabilities, especially in dynamic environments involving user mobility. This advantage equips the GNN to seamlessly adapt to evolving communication scenarios and maintain the communication rates under varying conditions.
\vspace{-0.26cm}
\subsection{Sample Complexity}
\begin{figure}[]
\centering
\includegraphics[width=0.36\textwidth]{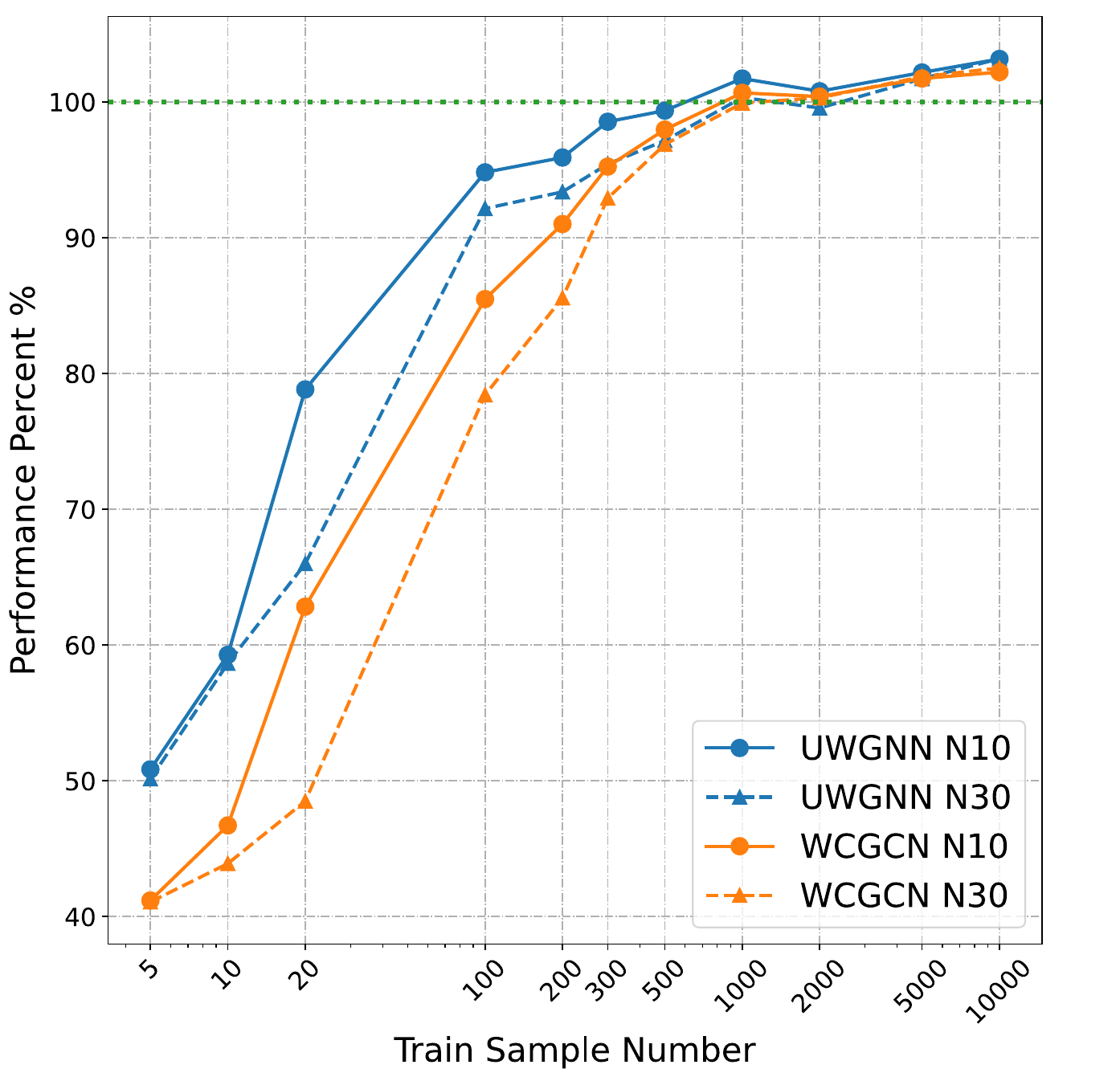}
\caption{Network sample complexity comparison}
\label{Fig.7}
\end{figure}
In \cite{r15}, the concept of network sample complexity and algorithm alignment is introduced, which shows that the higher algorithm alignment, the lower network sample complexity. To compare their network sample complexities, UWGNN and WCGCN are trained using different training samples until they attain convergence, and their performance is tested on 2000 test samples. Analysis of Fig. \ref{Fig.7} shows that our network performs well on small training datasets with different numbers of users, indicating that it does not depend on a large sample size to learn the statistical distribution of the data, but to learn the structure and iterative calculation of the WMMSE algorithm. Additionally, the performance improves faster with an increase in sample size, indicating that our network aligns more effectively with the algorithm than the conventional GNN network.

\section{Conclusion}
In this paper, we have proposed a novel knowledge-driven approach based on WMMSE algorithm inspired GNN to solve the resource allocation problem in D2D networks. Compared with current approaches, our approach has exhibited unique advantages in scalability and data generalization. Moreover, we have introduced a theoretical hypothesis for the validity of the graph neural network unrolling approach. Going forward, we plan to extend this work to other wireless resource allocation problems, such as bandwidth allocation, beam assignment, etc., and further develop our theoretical results to better guide the design of unrolling approaches. We expect that neural networks with knowledge-driven architecture will be significant in the future of wireless communication networks.

\bibliography{ref}
\bibliographystyle{IEEEtran}

\ifCLASSOPTIONcaptionsoff
 \newpage
\fi

\end{document}